\documentclass[aps,pra,reprint,twocolumn,10pt]{revtex4-2}

\usepackage{mathtools}
\usepackage{microtype}
\usepackage{graphicx}
\usepackage{siunitx}
\usepackage{dsfont}
\usepackage{braket}

\usepackage{booktabs}
\usepackage{hypernat}
     
\usepackage{caption}
\usepackage{subcaption}
\usepackage[svgnames]{xcolor}
\usepackage{hyperref}
\hypersetup{colorlinks,linkcolor={DarkBlue},citecolor={DarkBlue},urlcolor={DarkBlue}}
\usepackage[capitalise]{cleveref}

\newcommand{\e}{\operatorname{e}}

\begin{document}

\title{Mitigating linear optics imperfections via port allocation and compilation}
\author{Shreya P.~Kumar}
\affiliation{Xanadu, Toronto, ON, M5G 2C8, Canada}
\author{Leonhard Neuhaus}
\affiliation{Xanadu, Toronto, ON, M5G 2C8, Canada}
\author{Lukas G.~Helt}
\affiliation{Xanadu, Toronto, ON, M5G 2C8, Canada}
\author{Haoyu Qi}
\affiliation{Xanadu, Toronto, ON, M5G 2C8, Canada}
\author{Blair Morrison}
\affiliation{Xanadu, Toronto, ON, M5G 2C8, Canada}
\author{Dylan~H.~Mahler}
\affiliation{Xanadu, Toronto, ON, M5G 2C8, Canada}
\author{Ish Dhand}
\affiliation{Xanadu, Toronto, ON, M5G 2C8, Canada}
\date{\today}

\begin{abstract}
Linear optics is a promising route to building quantum technologies that operate at room temperature and can be manufactured scalably on integrated photonic platforms.
However, scaling up linear optics requires high-performance operation amid inevitable manufacturing imperfections.
We present techniques for enhancing the performance of linear optical interferometers by tailoring their port allocation and compilation to the on-chip imperfections, which can be determined beforehand by suitable calibration procedures that we introduce.
As representative examples, we demonstrate dramatic reductions in the average power consumption of a given interferometer or in the range of its power consumption values across all possible unitary transformations implemented on it.
Furthermore, we demonstrate the efficacy of these techniques at improving the fidelities of the desired transformations in the presence of fabrication defects.
By improving the performance of linear optical interferometers in relevant metrics by several orders of magnitude, these tools bring optical technologies closer to demonstrating true quantum advantage.
\end{abstract}

\maketitle

\section{Introduction}
Linear optics provides a scalable path to building a fault-tolerant quantum computer at room temperature~\cite{bourassa2021blueprint,Rudolph2017a}.
Boson sampling protocols and its variants based on linear optics are enabling striking demonstrations of quantum computational supremacy~\cite{Aaronson2013a,Zhong2020a,Hamilton2017,Kruse2019,deshpande2021quantum}.
Furthermore, linear optics is promising for obtaining an advantage in quantum communication~\cite{Duan2001a}, metrology~\cite{Motes2015a,Olson2017a} and in optical neural networks and neuro-morphic photonics~\cite{Shen2017,Steinbrecher2019,Harris2020}.

Rapid progress in integrated photonics, non-linear optics and photo-detection has led to an explosion in the size of linear-optical interferometers~\cite{capmany2020programmable,harris2018linear,taballione20188times,Zhong2020a}.
Operating these larger interferometers with desired levels of performance remains a challenge.
In particular, larger interferometers can be rendered ineffectual by tiny imperfections--- among them flaws related to power consumption and fabrication defects---that abate for smaller numbers of modes.

One challenge to scaling up interferometers is their high and variable power consumption.
Programmable interferometers, or photonic chips, consume power in order to implement different unitary transformations using reconfigurable phase-shifters.
Such phase-shifters are typically tuned by applying a voltage across an optical element based on thermo-optic or electro-optic phase manipulation~\cite{bogaerts2020programmable,van2010integrated,masood2013comparison,soref1987electrooptical,reed2010silicon,harris2017quantum,suzuki2018low,taballione202012}.
The application of this voltage consumes power that scales with the square of the number of modes in the chip.
Furthermore, implementing different transformations requires different phase-shifter settings~\cite{clements2016optimal,reck1994experimental,guise2018simple}, which leads to varying power consumption values and operating temperatures for different transformations implemented on the chip.
Large or varying power consumption values could rule out the promising direction of chip-integrated photodetectors: placing chips in cryogenic conditions is incompatible with large or varying heat loads~\cite{thiele2020cryogenic,eltes2020an-integrated}.
Furthermore, high and varying operating temperatures degrade the performance of the chip as the optical elements integrated on chip are often sensitive to temperature.
For instance, on-chip squeezers, which are important for Gaussian boson sampling and photonic quantum information processing, have a limited range of operating temperatures~\cite{vaidya2020broadband}.
Finally, while it is possible to correct for cross-talk through appropriate calibration and modelling~\cite{milanizadeh2020control}, minimizing the gradient through the use of a more constant power dissipation can reduce cross-talk effects~\cite{fatemi2019nonuniform}.
Thus it is important to devise methods to reduce the overall power consumption or to reduce the range of powers values required to implement different unitary transformations.

Another critical challenge is the reduction in fidelities of implemented transformations arising from fabrication defects.
Although existing methods can mitigate the effect of optical losses~\cite{su2020error} and discrete defects arising on individual optical elements~\cite{dhand2019circumventing}, 
there is no known method to mitigate imperfect splitting ratios of on-chip beam-splitters without incurring adverse overheads.
This imperfection is detrimental because balanced beam-splitters, in addition to reconfigurable phase-shifters, are required by current photonic chips in order to implement arbitrary unitary transformations~\cite{clements2016optimal}.
Fabrication imperfections can lead to imbalanced splitting ratios because these ratios depend on evanescent coupling or multi-mode interference, which are very sensitive to the dimensions and physical properties of the fabricated waveguides~\cite{crespi2013integrated,wang2019high}.
As we show below, this imperfection is relatively harmless for small interferometers but can quickly degrade the fidelities of the implemented transformations, rendering larger chips inaccurate even with small fabrication defects.
While adding complementary reconfigurable beam-splitters can help compensate for fabrication imperfections~\cite{suzuki2015ultra-high-extinction-ratio},they introduce many new optical and electrical elements.
Alternatively, nonlinear optimization can be used to improve the choice of interferometer phases,~\cite{mower2015high-fidelity} but this method 
suffers from optimization convergence issues and high classical computational overheads, especially for large interferometers that stand to gain most from such an approach.

We present a new means of mitigating these and other challenges that might arise on current and future photonic chips. Our versatile tool-kit works by changing how the chips are operated rather than how they are designed.
In particular, we introduce the method of port allocation, which exploits the hitherto unused degree of freedom of labelling the input and output ports of an interferometer.
We show that port allocation is effective at reducing the total power consumption of interferometers, and it can reduce power ranges by several orders of magnitude, thus enabling near-constant chip operating temperatures. 
We also introduce a hardware-tailored compiler that reduces the effect of imperfect splitting ratios by several orders of magnitude.
The splitting ratios required by this compiler can be determined accurately by suitable calibration procedures that we devise.
Finally, we show that the combination of the calibration, tailored compilation and port-allocation provides dramatic improvements in the performance of chips in the presence of fabrication defects.

\section{Background}
Linear optics refers to the set of optical transformations generated by Hamiltonians that are linear in the electric and magnetic fields of light.
Discrete linear-optical transformations acting on $N$ modes of light are described by the $N\times N$ unitary matrix $U$, which connects the input bosonic operators $\{a_{k}: k \in 1, 2, \dots, N\}$ to the output operators according to $a_{j} \mapsto \mathcal{U} a_{j} \mathcal{U}^{\dag} = \sum_{k=1}^{N} U_{kj} a_{k}$.
The operator $\mathcal{U}$ describes the transformation on the full state of light while the $N\times N$ matrix $U$ provides a compact representation of this transformation~\cite{book_serafini,kok2007linear}.
Henceforth operators (matrix representations) are denoted by a calligraphic font (Roman font).

Arbitrary $N\times N$ unitary matrices $U$ can be implemented in the spatial domain using $N$-mode interferometers. 
Integrated (on-chip) interferometers require $U$ to be decomposed into smaller two-mode unitary matrices that can be directly implemented using optical elements such as beam-splitters and phase-shifters~\cite{clements2016optimal,reck1994experimental,guise2018simple}.
The standard implementation of these two-mode unitary matrices is via tunable Mach-Zehnder interferometers (MZI), which comprise two static balanced beam-splitters (i.e., with 50:50 splitting ratios) placed in alternation with tunable phase-shifters according to
\begin{equation}
\begin{split}
\includegraphics[width = 0.6\columnwidth]{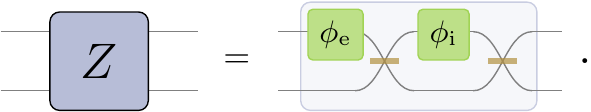}
\label{Eq:M}
\end{split}
\end{equation}
Here, the MZI matrix $Z$ depends on the phase-shifter parameters $\phi_\text{e}$ and  $\phi_\text{i}$ as
\begin{equation}
Z (\phi_{\text{i}}, \phi_{\text{e}}) = \, i \e^{i \phi_{\text{i}}/2}
\begin{pmatrix*}[c]
\e^{i \phi_{\text{e}}} \, \sin{\phi_{\text{i}}/2} & \cos{\phi_{\text{i}}/2} \\
\e^{i \phi_{\text{e}}} \, \cos{\phi_{\text{i}}/2} & - \sin{\phi_{\text{i}}/2}
\end{pmatrix*},
\label{Eq:MZ_Clements_main}
\end{equation}
where subscripts i and e refer to internal and external phase-shifters.

Linear-optical decompositions~\cite{clements2016optimal,guise2018simple,reck1994experimental} map the desired unitary operator to the phases $\{\phi_{\text{i}}\}$ and $\{\phi_{\text{e}}\}$ of the MZIs networked on the chip.
Although the algorithm to find the phases from the given unitary matrix is straightforward, the actual relation between the elements of the unitary matrix and the phases is rather complicated: changing one or a few of the elements of the unitary matrix generally results in completely different sets of phases required to implement the transformation.

Reconfigurable phase-shifters on photonic chips are typically implemented using thermo-optic or electro-optic phase manipulation, which use heat or current to change the waveguide properties~\cite{bogaerts2020programmable,van2010integrated,masood2013comparison,soref1987electrooptical,reed2010silicon}.
In particular, large scale interferometric networks often use the thermo-optic effect as it allows for compact implementations on versatile platforms~\cite{harris2017quantum,suzuki2018low,taballione202012}.
These effects lead to the consumption of electrical power by the phase-shifters, whose amount is determined by the phases to be implemented.
The phase-voltage relation for each phase-shifter is different: $\phi_{\ell} = \alpha_{\ell} + \beta_{\ell} V_{\ell}^{2}$, where the index $\ell$ runs over all internal and external phase-shifters. The offset $\alpha_{\ell}$ and slope $\beta_{\ell}$---which can be vastly different for each phase shifter---can be determined during the calibration of the chip~\cite{arrazola2021quantum}.
Although here we assume thermal cross-talk to be negligible, the analysis of this work is also applicable to the case of non-negligible cross-talk.

Thus, each unitary operator $U$ is eventually implemented via voltages $\left\{ V_{\ell}\right\}$ that are applied to the phase-shifters.
Furthermore, the total power consumed is $\propto\sum_{\ell} V_{\ell}^{2}$.
Thus, the power required to implement a unitary matrix can be calculated according to
\begin{equation}
	U \stackrel{\text{decompose}}{\mapsto} \left\{\phi_{\ell} \right\} \stackrel{\text{calibrate}}{\mapsto} \left\{ V_{\ell}\right\}  \stackrel{\text{square, sum}}{\mapsto} \text{Power}.
\end{equation}
In other words, different unitary transformations consume different amounts of power, which can lead to different chip operating temperatures.

\section{Error mitigation via port allocation}
\begin{figure}
\includegraphics[width = \columnwidth]{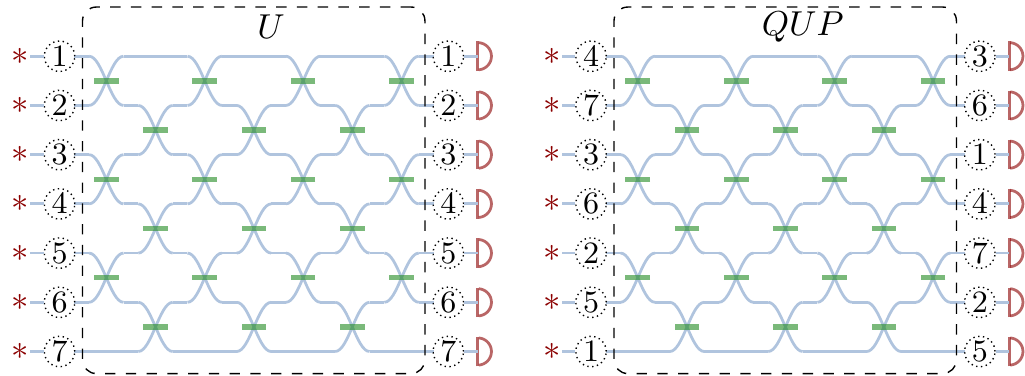}
 \captionof{figure}{
 	\textbf{Port allocation in linear optics.}
	Labels of input and output ports of two interferometers performing the same computation under default allocation (left) and after allocating to mitigate imperfections (right).
 	The column of numbers on the left represent the allocation of the sources ($*$ sign) and input ports, and those on the right represent the allocation of output ports and detectors (closed semi-circle), respectively.
 	The measurement outcomes are ideally unchanged under the permutation of the source and input labels, and of the output and detector labels, if accompanied by suitable permutation $QUP$ of the unitary matrix $U$ implemented by the interferometer.
 }
 \label{Fig:Relabeling}
 \end{figure}

Chip imperfections can be mitigated by exploiting the additional degree of freedom provided by the labels of the input and output ports.
Our method of port allocation determines the permutation of these labels that minimizes a chosen imperfection.
For an ideal interferometer, different port allocations would return the same measurement outcomes if the input light and the detectors at the output were also labelled suitably, as depicted in \cref{Fig:Relabeling}.
However, for imperfect chips, different port allocations can lead to vastly better performance, which is the main insight behind this method.

In more detail, consider light in state $\ket{\psi}$ incident at an $N$-mode interferometer $\mathcal{U}$ and measured at detectors described by operators $\sum_m \mathcal{M}_{m}^{\dagger}\mathcal{M}_{m} = \mathds{1}$~\cite{nielsen2002quantum}.
We now show that permuting the input and output ports of the interferometer leaves the probabilities unchanged if the incident light and the measurement operators are suitably permuted as well.
The probability of obtaining measurement outcome $m$ is $p(m) = \braket{\psi | \mathcal{U}^{\dag} \mathcal{M}_{m}^{\dag} \mathcal{M}_{m} \mathcal{U} | \psi}$.
This probability is unchanged if suitable permutation operators and their inverse $\mathcal{P}\mathcal{P}^{-1}$ and $\mathcal{Q}\mathcal{Q}^{-1}$ are applied before and after the interferometer.
That is, $p(m) = \tilde{p}(m)$ for $\tilde{p}(m) := \braket{\widetilde{\psi} | \widetilde{\mathcal{U}}^{\dag} \widetilde{\mathcal{M}}_{m}^{\dag} \widetilde{\mathcal{M}}_{m} \widetilde{\mathcal{U}} | \widetilde{\psi}}$, where $\widetilde{\mathcal{U}} = \mathcal{Q} \mathcal{U} \mathcal{P}$, $\widetilde{\mathcal{M}}_{m} = \mathcal{Q} \mathcal{M}_{m} \mathcal{Q}^{-1}$ and $\ket{\widetilde{\psi}} = \mathcal{P}^{-1} \ket{\psi}$.

Each of these operators $\{\mathcal{O} \in \mathcal{U}, \mathcal{P}, \mathcal{Q}\}$ possess compact matrix representation $\{O \in U, P, Q\}$ with $\mathcal{O} a_{j} \mathcal{O}^{\dag} = \sum_{k} O_{kj} a_{k}$.
For permutation operators, we moreover define permutation functions $\pi(j)$ and $\sigma(j)$ such that $\mathcal{P} a_{j} \mathcal{P}^{-1} = \sum_{k} P_{kj} a_{k}  =: a_{\pi(j)}$ and $\mathcal{Q} a_{j} \mathcal{Q}^{-1} = \sum_{k} Q_{kj} a_{k} =: a_{\sigma(j)}$.

From this relation between the operators and their matrix representations, we obtain the matrix relation of the permuted interferometer, i.e., $\widetilde{\mathcal{U}} a_{j} \widetilde{\mathcal{U}}^{\dag} = \sum_{k} \widetilde{U}_{kj} a_{k}$ for $\widetilde{U} = Q U P$. 
Thus, by permuting the input state and the measurements at the output, we can perform the same experiment but with a different, permuted interferometer $\widetilde{U}$.
This permuted interferometer can have completely different response to imperfections, e.g., different power consumption and different fidelity in the presence of fabrication defects (see Appendix \cref{Fig:PowerChange}).

\begin{figure}
\includegraphics[width = \columnwidth]{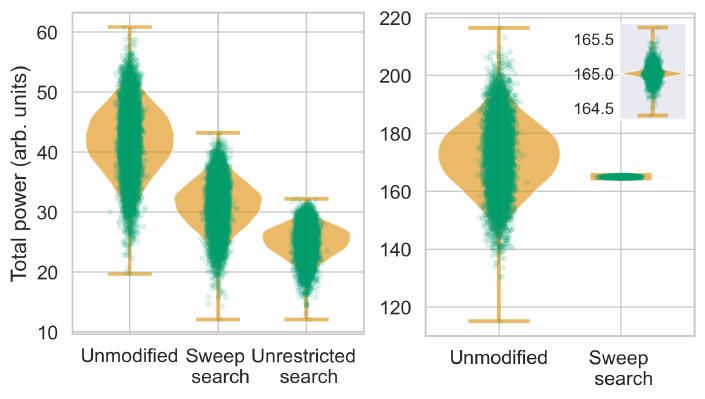}
\caption{\textbf{Port allocation for minimizing overall power consumption (left) and range (right) -- violin and scatter plots with jitter.}
(left) Total power consumed by a 4-mode chip for \num{10000} Haar-random unitary matrices implemented on a chip with random phase-voltage relations using different algorithms for finding suitable port allocations: unrestricted search and Sweep Search ($k=2$).
(right) A target power is specified, 165 units in this case, and the Sweep Search algorithm ($k=2$) is used to find allocations that minimize the distance from the target. 
The total power consumed by an 8-mode interferometer to implement \num{10000} Haar-random unitaries is depicted.
Inset: close-up of the distribution after port allocation.
Suitable port allocation can reduce the total and range of powers consumed by the interferometers.
\label{Fig:AllocationPower}}
\end{figure}

The analysis simplifies if the input is a product state and when identical measurements are performed independently on the output ports, as is true for the important cases of Gaussian and Aaronson-Arkhipov boson sampling~\cite{Aaronson2013a,Zhong2020a,Hamilton2017,Kruse2019,deshpande2021quantum}. 
The permutation operators, $\mathcal{P}$ and $\mathcal{Q}$, transform any function of $a_{i}$ and $a^{\dag}_{j}$ according to 
$\mathcal{P} f (a_{i}, a^{\dag}_{j} ) \mathcal{P}^{-1} = f (a_{\pi(i)}, a^{\dag}_{\pi(j)})$ and $\mathcal{Q} g (a_{i}, a^{\dag}_{j} ) \mathcal{Q}^{-1} = g (a_{\sigma(i)}, a^{\dag}_{\sigma(j)} )$.
If the input is in a product state $\ket{\psi} = \otimes_{i = 1}^{N} f (a_{i}, a^{\dag}_{i} ) \ket{0}$, then the permuted input state $\ket{\widetilde{\psi}} = \mathcal{P}^{-1} \ket{\psi}$ is also a product state.
That is, $\ket{\widetilde{\psi}} = \otimes_{i = 1}^{N} f \left(a_{\pi^{-1}(i)}, a^{\dag}_{\pi^{-1}(i)} \right) \ket{0}$.
As an example, if the squeezing values of the input states of a Gaussian boson sampling device are $\{s_{1}, s_{2}, \dots, s_{N} \}$, then the permuted squeezing values are  $\{s_{\pi^{-1}(1)}, s_{\pi^{-1}(2)}, \dots, s_{\pi^{-1}(N)} \}$.
Similarly, if the measurements possess a tensor product structure, i.e., are non-entangling, then the measurement outcomes transform from $\{m_{1}, m_{2}, \dots, m_{N} \}$, to  $\{m_{\sigma(1)}, m_{\sigma(2)}, \dots, m_{\sigma(N)}\}$.
Thus, for such problems, the permutation can be performed at the software level alone, i.e., the input squeezing values are permuted according to $\pi^{-1}$, the programmed interferometer is changed to $QUP$, and the detected photon pattern is permuted according to $\sigma$ before being returned to the user.
This is depicted in \cref{Fig:Relabeling}.

Thus, permuting the input and output ports of an interferometer can be used to change its properties, and searching over these different allocations allows for mitigating imperfections.
There are a total of $n!^2$ unique allocations of the input and output ports, which provides access to a large space of port allocations with reduced imperfections.
This large search space also necessitates developing efficient procedures to search over the different allocations.
An unrestricted search over the different permutations might be feasible for a few (say 3--8) modes but becomes prohibitively expensive for larger interferometers.
We devise a faster search algorithm, Sweep Search, that substantially narrows the search space without compromising significantly on performance. 
The idea behind Sweep Search is to identify a small set of port permutations as being most impactful at changing the interferometer phases and to search over these permutations and their compositions (See \cref{Sec:PPMethods}).
Sweep Search scales as $k 2^n$ and performs suitably for even small values of constant parameter $k \sim 2$.

\cref{Fig:AllocationPower} depicts the performance of the unrestricted search and the Sweep Search algorithms at reducing power consumption.
While unrestricted search reduces the mean power consumption of the interferometer over unitary matrices chosen from the Haar measure by over 40\%, the significantly faster Sweep Search leads to over 25\% reduction in power.
The figure also presents the reduction in the variability of the power consumption using the Sweep Search algorithm.
A remarkable factor of $139.3$ reduction in the standard deviation of the power-consumption distribution is observed.
We simulated over a hundred different phase-voltage relations and always saw similar improvements. \cref{Fig:AllocationPower} is a representative plot for one such interferometer with random phase-voltage relations."

\section{Error mitigation via hardware tailored decomposition}
We now present a method to deal with the effects of fabrication defects, focussing on imperfect splitting ratios of the beam-splitters on the chip.
Consider first the effect of this imperfection on a single MZ interferometer. 
If the reflectivity of an individual beam-splitter is  $\cos^2\theta$, then MZ interferometers implement the transformation $Z^{\theta}(\phi_{\text{i}},\phi_{\text{e}})=$
\begin{equation}
    \begin{pmatrix*}[c]
    \e^{i \phi_{\text{e}}} \left(\e^{i \phi_{\text{i}}} \cos^{2}{\theta} - \sin^{2}{\theta}\right)  & i \e^{i \frac{\phi_{\text{i}}}{2} } \cos\frac{\phi_{\text{i}}}{2} \sin{2 \theta}\\
     i \e^{i \left( \phi_{\text{e}} + \frac{\phi_{\text{i}}}{2} \right)}  \cos \frac{\phi_{\text{i}}}{2} \sin{2 \theta} & - \e^{i \phi_{\text{i}}} \sin^{2}{\theta} + \cos^{2}{\theta}
    \end{pmatrix*}
\label{Eq:MZ_Trusplit_main}
\end{equation}
instead of transformation~\eqref{Eq:MZ_Clements_main}.
Note that $Z^{\frac{\pi}{4}} = Z$.
This imperfection restricts the range of reflectivity values that an MZ interferometer can implement: while the reflectivity of the perfect MZI~\eqref{Eq:MZ_Clements_main} is $\sin^{2} (\phi_{\text{i}}/2) $, which can range from 0 to 1, the reflectivity of the imperfect MZI \eqref{Eq:MZ_Trusplit_main} is $1 - \sin^{2} (2 \theta) \cos^{2} (\phi_{\text{i}}/2) $, which ranges from $\cos^{2} (2 \theta)$ to $1$.

With current decomposition procedures~\cite{clements2016optimal,guise2018simple,reck1994experimental}, the phases $\phi_{\text{i}}, \phi_{\text{e}}$ are calculated for a given target unitary $U_{\text{t}}$ assuming that $\theta = \pi/4$.
The resulting unitary matrix $U_{\text{r}}$ implemented on the chip is actually a composition of $Z^{\theta}(\phi_{\text{i}},\phi_{\text{e}})$ transformations instead of $Z(\phi_{\text{i}},\phi_{\text{e}})$ transformations.
This difference leads to inaccurate implementation of the desired unitary transformation, and the inaccuracy increases with $|\theta - \pi/4|$ and with the size of the unitary.

\begin{figure}
\includegraphics[width = \columnwidth]{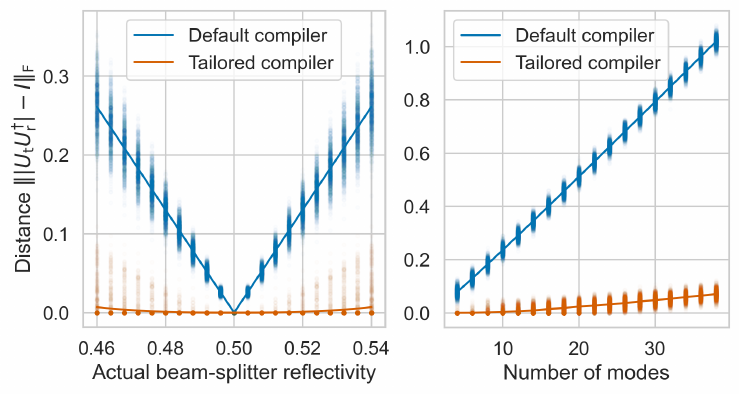}
\caption{\label{Fig:Trusplit}
\textbf{Improvement in accuracy of implemented unitary using tailored compilation.}
Distance $\Vert |U_{\mathrm{t}} U_{\mathrm{r}}^{\dagger}| - I \Vert_\text{F}$ (where $\Vert\bullet\Vert_{\text{F}}$ denotes the Frobenius norm) due to beam-splitter imperfections between \num{1000} implemented $U_{\mathrm{r}}$ and target unitaries $U_{\mathrm{t}}$ drawn from the Haar measure. 
(left) A scatter plot of the distance as a function of beam-splitter reflectivity for a 6-mode chip; solid lines represent the mean distances over the different target unitaries.
(right) Scatter plot and mean values of the distance as a function of number of modes on the chip for a splitting ratio of $48$:$52$.}
\end{figure}

We propose to instead decompose $U_{\text{t}}$ directly by \textit{nulling}~\cite{clements2016optimal} using \cref{Eq:MZ_Trusplit_main} instead of \cref{Eq:MZ_Clements_main}.
The steps in the decomposition are similar to the original decompositions, 
i.e., different elements of the unitary matrix $U$ are nulled in a specific order using appropriate choices of MZ matrices $Z^{(\theta)}$.
However, decomposing an arbitrary unitary into imperfect MZ interferometers is not possible in general: there might not be a value of  $\phi_\text{i}$ that will null the desired matrix element.
In such cases, the phase is set to either $2 \pi$ or $0$ to achieve the closest possible MZI behaviour.
The new nulling equations and a comparison with the the procedure of Ref.~\cite{clements2016optimal} are presented in~\cref{Tab:Comparision} of \cref{Sec:Trusplit}.
Our procedure can also be used if each MZI has a different splitting ratio by setting a suitable value of $\theta$ for each nulling MZI matrix.

This tailored decomposition enhances the fidelities of the implemented unitaries by many order of magnitude, as depicted in \cref{Fig:Trusplit}. 
A factor 550 reduction in mean distance is observed for a reasonable splitting ratio of $49$:$51$ for a 6-mode chip.
A stark enhancement is observed for chips with tens of modes, where the tailored compilation helps avoid the large distance values that could be detrimental for the accurate operation of the chip. 

\begin{figure}
\includegraphics[width = \columnwidth]{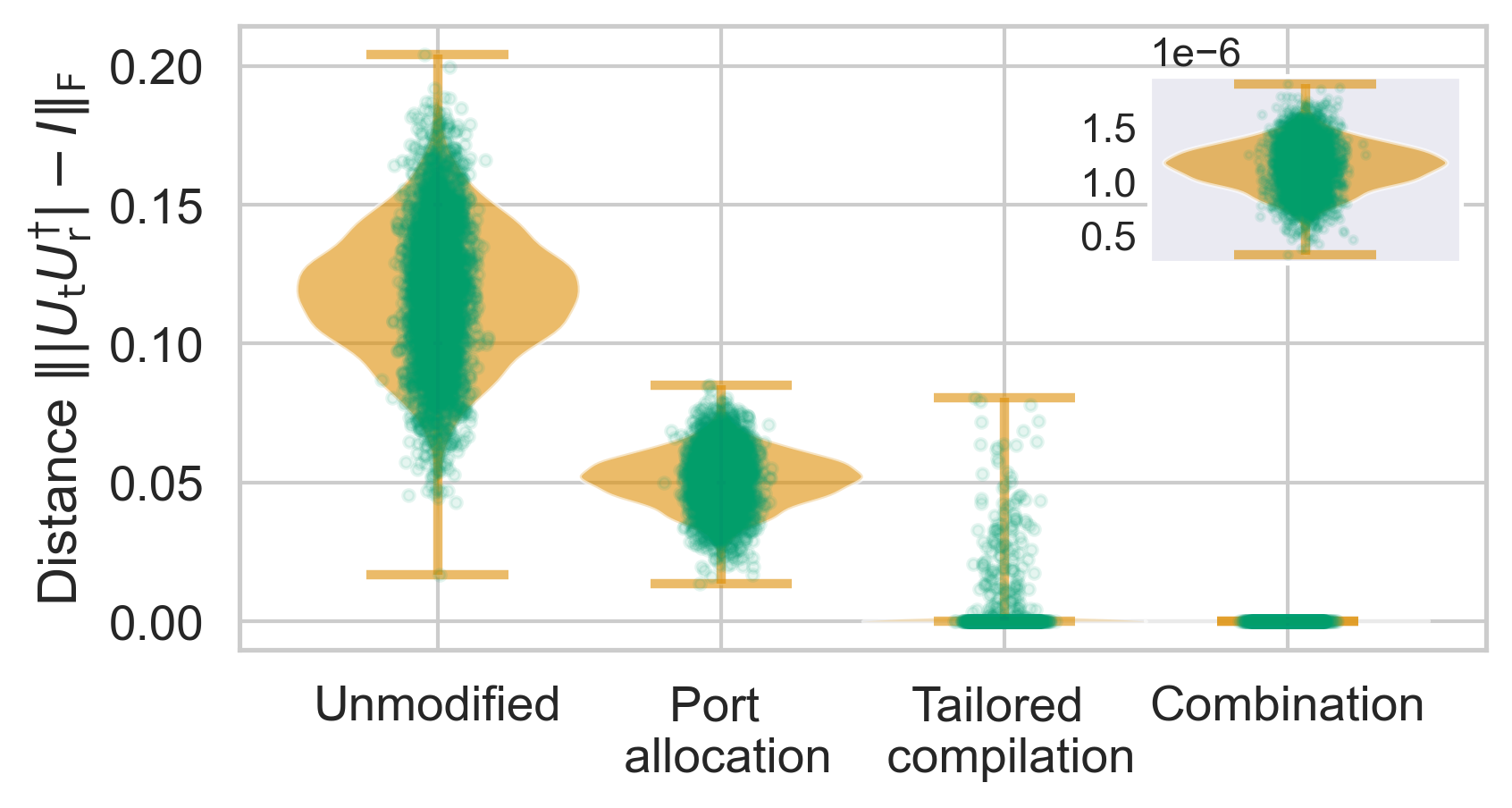}
\caption{\label{Fig:Pipeline}
\textbf{Simulation of complete work-flow to mitigate imperfect splitting ratios}.
The beam-splitter reflectivity values of a 4-mode chip are equal to $0.47$ but unknown to a user. \num{5000} unitaries are drawn from the Haar measure and an estimate of the splitting ratio is determined using the procedure described in the main text, and port allocation and tailored compilation are applied individually and in combination to improve the accuracy of the implemented transformation.
Inset: a close-up of the distribution after both methods are used in combination.
While tailored compilation can improve performance for most unitaries, there are still a few outliers remaining that have low fidelity, and a combination of both methods removes these outliers.
}
\end{figure}

Consider now the task of determining the values of the reflectivity parameters $\theta$ of the MZIs on a given chip.
One method to find these relies on identifying that imperfect MZIs \eqref{Eq:MZ_Trusplit_main} can implement perfectly reflective transformations when $\phi_{\text{i}} = \pi$, but perfect transmission is not possible as the reflectivity is bounded from below by $\cos^{2} (2 \theta)$ for $\phi_{\text{i}} = 0$.
Thus, by sequentially setting each MZI to $\phi_{\text{i}} = 0$ while setting the rest of the MZIs to $\phi_{\text{i}} = \pi$, the $\theta$ value of each of the MZIs can be recovered from intensity measurements.
Another method operates under the assumption that the $\theta$ values are uniform across the chip; this model is reasonable if this imperfection arises from systematic defects in fabrication but this single value captures  the imperfection well even if the assumption is not met (\cref{Fig:PipelineErrors} in \cref{Sec:NonUniform}).
In this method, a specific pre-determined transformation is implemented on the chip with the tailored decomposition assuming an initial value of $\theta = \theta_{0}$ and the difference between the expected and observed intensities is noted.
The procedure then iteratively chooses different values of $\theta$ that optimize this distance and concludes when the distance converges to a (local) minimum.
\cref{Sec:Calibration} details the two procedures.

In \cref{Fig:Pipeline}, we combine the different methods presented above, namely the procedure to determine the imperfect splitting ratio and port allocation in conjunction with the tailored decomposition. 
We consider a situation where the splitting ratios are equal but unknown to a user.
The second calibration procedure is used to determine $\theta$, and this value is used in the tailored compiler.
The combination of port allocation and tailored compilation provides an enhancement of $5$ orders of magnitude in the distance between the target and the resulting unitary.
An enhancement is also seen when the splitting ratios are all different but are assumed to be the same (\cref{Fig:PipelineErrors} in \cref{Sec:NonUniform}).

\section{Conclusion}%
We have presented methods to mitigate various imperfections on photonic chips.
The method of port allocation demonstrates a remarkable improvement in the power consumption properties of a chip, while the tailored compilation procedure and especially a combination of two methods mitigate fabrication defects.
Thus, these procedures can be used to address a variety of imperfections on current and future chips, enable high-performance operation of photonic chips and provide a significant boost to photonic quantum information processing.

\section*{Acknowledgements}
The authors are grateful to Ilan Tzitrin and Nicol\'as Quesada for helpful discussions.

\appendix

\section{Implementing linear optics with beam-splitters and phase-shifters}
\label{Sec:Background}

In this appendix, we provide relevant background on implementing linear optics using beam-splitters and phase-shifters
In particular, we recap decomposition procedures used for implemented arbitrary linear optical transformations using MZIs, which are in-turn implemented using balanced beam-splitters and tunable phase-shifters.

\subsection{Decomposition of unitary matrix into MZI matrices}
 Standard decomposition algorithms~\cite{clements2016optimal,guise2018simple,reck1994experimental} decompose an $N$-mode unitary into $Z_{[mn]}(\phi_{\text{i}}, \phi_{\text{e}})$ matrices and its inverse $Z_{[mn]}^{-1}(\phi_{\text{i}}, \phi_{\text{e}})$, which are $N \times N$ identity matrices whose $m$th row and $n$th column are set to the elements of the matrix $Z$ (Eq.~\ref{Eq:MZ_Clements_main}) and its inverse $Z^{-1}$.

For a detailed review of different decompositions, see Appendix~B of Ref.~\cite{kumar2021unitary}.
Here we recall the basic principle behind the most common decomposition algorithms.
The decomposition algorithms rely on sequentially nulling elements of the unitary matrix $U$ by multiplying it with $Z_{[mn]}(\phi_{\text{i}}, \phi_{\text{e}})$ or $Z_{[mn]}^{-1}(\phi_{\text{i}}, \phi_{\text{e}})$ and choosing values of $\phi_{\text{i}}$ and $\phi_{\text{e}}$ such that the elements are nulled.
This nulling is carried out until only a diagonal matrix remains.
Inverting the relation yields the decomposition of the matrix $U$.
In more mathematical terms, if we can choose the right sequence of matrices to null elements of $U$ until we have a diagonal matrix $D$
\begin{equation}
Z_{[m_{1}n_{1}]} Z_{[m_{2}n_{2}]} \dots U Z^{-1}_{[m_{3}n_{3}]} Z^{-1}_{[m_{4}n_{4}]} \dots = D,
\label{Eq:ClementsNulling}
\end{equation}
then we obtain the decomposition of $U$ as
\begin{equation}
U =  \dots Z^{-1}_{[m_{2}n_{2}]} Z^{-1}_{[m_{1}n_{1}]} D  \dots  Z_{[m_{4}n_{4}]}  Z_{[m_{3}n_{3}]}.
\label{Eq:Clements1}
\end{equation}
Since $D$ is a diagonal matrix of phase elements, we can always find matrices $D'$ and phases $\phi'_{\text{i}}, \phi'_{\text{e}}$ such that
\begin{equation}
Z^{-1}_{[mn]}(\phi_{\text{i}}, \phi_{\text{e}}) D = D' Z_{[mn]}(\phi'_{\text{i}}, \phi'_{\text{e}}).
\label{Eq:ClementsPhaseEnd}
\end{equation}
By using this relation repeatedly, $U$ can be decomposed into a product of matrices of the form $Z_{[mn]}$ followed by diagonal  matrix $D'$ acting on all the modes, bringing \cref{Eq:Clements1} into the form
\begin{equation}
U =  D' Z_{[m_{1}n_{1}]}  Z_{[m_{2}n_{2}]} \dots .
\label{Eq:Clements2}
\end{equation}
The phases $\phi_{\text{i}}$ and $\phi_{\text{e}}$ that comprise the $Z_{[mn]}$ matrices are set on the interferometer to implement the unitary $U$.

Different sequences of initial nulling orders (i.e., before the phases are moved to the left) lead to different geometries of the decomposed smaller unitaries.
For example, nulling can be performed only using $Z_{[mn]}(\phi_{\text{i}}, \phi_{\text{e}})$ matrices multiplied to the left, which results in a circuit that is triangular in structure proposed by Reck~\textit{et al.}~\cite{reck1994experimental} and a similar more recent structure proposed by de Guise \textit{et al.}~\cite{guise2018simple}.
Another possible decomposition, where initial nulling is performed by multiplying $Z_{[mn]}(\phi_{\text{i}}, \phi_{\text{e}})$ matrices to the left and $Z_{[mn]}^{-1}(\phi_{\text{i}}, \phi_{\text{e}})$ matrices to the right, leads to a rectangular circuit structure of Clements~\textit{et al.}~\cite{clements2016optimal}.

\subsection{Implementing MZI matrices with beam-splitters and phase-shifters}
These two-mode unitary matrices are implemented by tunable MZIs, which are comprised of static balanced (i.e., with 50:50 splitting ratios) beam-splitters and tunable phase-shifters as shown in \cref{Eq:M}.
The transformation implemented by the balanced beam-splitters on the interferometer is
\begin{align}
B_{\text{bal}} = \frac{1}{\sqrt{2}}
 \begin{pmatrix*}[c]
 1 & i \\
 i & 1
 \end{pmatrix*}.
\label{Eq:B_bal}
\end{align}
A phase-shifter implements the transformation
\begin{align}
R(\phi) =
	\begin{pmatrix*}[c]
	\e^{i\phi} & 0 \\
	0 & 1
	\end{pmatrix*}
\label{Eq:Phase}
\end{align}
for any tunable $\phi$.
Sandwiching a phase-shifter $R(\phi_{\text{i}})$ in between two balanced beam-splitters and by implementing an additional external phase $R(\phi_{\text{e}})$, we obtain the MZI matrix $Z$ given by
\begin{align}
Z(\phi_{\text{i}}, \phi_{\text{e}}) =& B_{\text{bal}} R(\phi_{\text{i}}) B_{\text{bal}} R(\phi_{\text{e}}) \\
=& \, i \e^{i \phi_{\text{i}}/2}
\begin{pmatrix*}[c]
\e^{i \phi_{\text{e}}} \, \sin{\phi_{\text{i}}/2} & \cos{\phi_{\text{i}}/2} \\
\e^{i \phi_{\text{e}}} \, \cos{\phi_{\text{i}}/2} & - \sin{\phi_{\text{i}}/2}
\end{pmatrix*}.
\label{Eq:MZ_Clements}
\end{align}

By tuning $\phi_{\text{i}}$ and $\phi_{\text{e}}$, we can implement any $2\times 2$ special unitary $\text{SU}(2)$ matrix.
However, we note that this representation is not unique.
For example, the $T$-matrices considered in Ref.~\cite{clements2016optimal} are given by
\begin{align}
T_{1}(\phi_{\text{i}}, \phi_{\text{e}})  =
	\begin{pmatrix*}[c]
	\e^{i\phi_{\text{e}}}\,\cos{\phi_{\text{i}}} & -\sin{\phi_{\text{i}}} \\
	\e^{i\phi_{\text{e}}}\,\sin{\phi_{\text{i}}} & \cos{\phi_{\text{i}}}
	\end{pmatrix*},
\end{align}
and those considered in Ref.~\cite{reck1994experimental} have the form
\begin{align}
T_{2}(\phi_{\text{i}}, \phi_{\text{e}})  =
	\begin{pmatrix*}[c]
	\e^{i\phi_{\text{e}}}\,\sin{\phi_{\text{i}}} & \e^{i\phi_{\text{e}}} \,\cos{\phi_{\text{i}}} \\
	\cos{\phi_{\text{i}}} & -\sin{\phi_{\text{i}}}
	\end{pmatrix*}.
\end{align}
This is only a different definition of the angles and convention of phases.
However, we adopt the convention of \cref{Eq:MZ_Clements} in these calculations because it represents one standard hardware realization of on-chip MZIs.

\section{GBS as an example of port allocation}
\label{Sec:GBSExample}
In this section, we provide an illustrative example of the port-allocation procedure for a Gaussian boson sampling (GBS) circuit~\cite{Hamilton2017,Kruse2019}.
The special feature of GBS that makes this example especially appealing is that both the input light and the measurement operators have a tensor-product form.
Hence, as described in the main text, the port-allocation can be performed at the software level alone by merely changing the programming of the inputs, unitary and measurement outcomes.

Let us focus on a four-mode circuit implementing a GBS problem and the application of reducing its power consumption mean or range.
GBS circuits comprise single-mode squeezed states of light incident at each of the ports of an arbitrary linear optical circuit to generate a Gaussian state, which is then measured at photon number resolving detectors. 
A pattern of clicks at the detectors is the desired GBS sample.
The input received from the user comprises: firstly a $4 \times 4$ unitary matrix, say $U$ and, secondly a set of four squeezing values, say $\{s_{1}, s_{2}, s_{3}, s_{4}\}$, specifying the state of the light impinged at the four input ports in order.
Before the experiment is run, the phases required to implement $U$ are calculated using a chosen decomposition of the unitary matrix into beam-splitters~\cite{clements2016optimal,guise2018simple,reck1994experimental} or into larger optical circuits \cite{Dhand2015a,Su2019a,kumar2021unitary}.
These phases are then used to calculate the voltages  $\{V_{1}, V_{2}, \dots,\}$ that are required to implement the transformation $U$.
The total power consumption of this interferometer is proportional to $\sum_{\ell} V_{\ell}^{2}$
The exact relation between phases and voltages can be determined with a calibration procedure run on the interferometer.
The light emitted from the interferometer is measured at photon-number resolving (PNR) detectors, and say $\{n_{1}, n_{2}, n_{3}, n_{4}\}$ photons respectively are recorded as being emitted from the four output ports.

To mitigate imperfections on the interferometer, an input-port permutation (say $P$) and and output-port permutation (say $Q$) are chosen.
The exact choice of the permutations depends on the relevant application and can be chosen by suitable algorithms as described in the next section.
The permutations can be considered to be in their compact matrix representation as discussed in the main text.
Multiplying a permutation matrix from matrix to the left of another matrix leads to a reordering of the rows of the other matrix, while right multiplication corresponds to reordering of the columns.

The new unitary transformation to be performed on the interferometer is $\widetilde{U} = Q U P$.
The action of these permutation matrices leads to the rows and columns of $U$ being shuffled.
For instance, consider an input permutation of the form
\begin{equation}
	P =
	\begin{pmatrix}
	0 & 1 & 0 & 0\\
	0 & 0 & 1 & 0\\
	1 & 0 & 0 & 0\\
	0 & 0 & 0 & 1
	\end{pmatrix},
\end{equation}
which cycles through the first three indices and leaves the fourth index unchanged.
Furthermore, assume that the output permutation is
\begin{equation}
	Q =
	\begin{pmatrix}
	1  & 0 & 0 & 0\\
	0 & 1 & 0 & 0\\
	0 & 0 & 0 & 1\\
	0 & 0 & 1 & 0
	\end{pmatrix},
\end{equation}
which leaves the first two indices unchanged and swaps the last two indices.
The actions of these permutations on the original unitary matrix
\begin{equation}
	U =
	\begin{pmatrix}
	U_{11} & U_{12} & U_{13} & U_{14}\\
	U_{21} & U_{22} & U_{23} & U_{24}\\
	U_{31} & U_{32} & U_{33} & U_{34}\\
	U_{41} & U_{42} & U_{43} & U_{44}
	\end{pmatrix},
\end{equation}
will shuffle the rows and columns to give new unitary matrix
\begin{equation}
	\widetilde{U} = QUP =
	\begin{pmatrix}
	U_{12} & U_{13} & U_{11} & U_{14}\\
	U_{22} & U_{23} & U_{21} & U_{24}\\
	U_{42} & U_{43} & U_{41} & U_{44}\\
	U_{32} & U_{33} & U_{31} & U_{34}
	\end{pmatrix}.
\end{equation}
Because permutations are also unitary transformations, we see that the permuted matrix $\widetilde{U}$ is also unitary.

\begin{figure}
\includegraphics[width = \columnwidth]{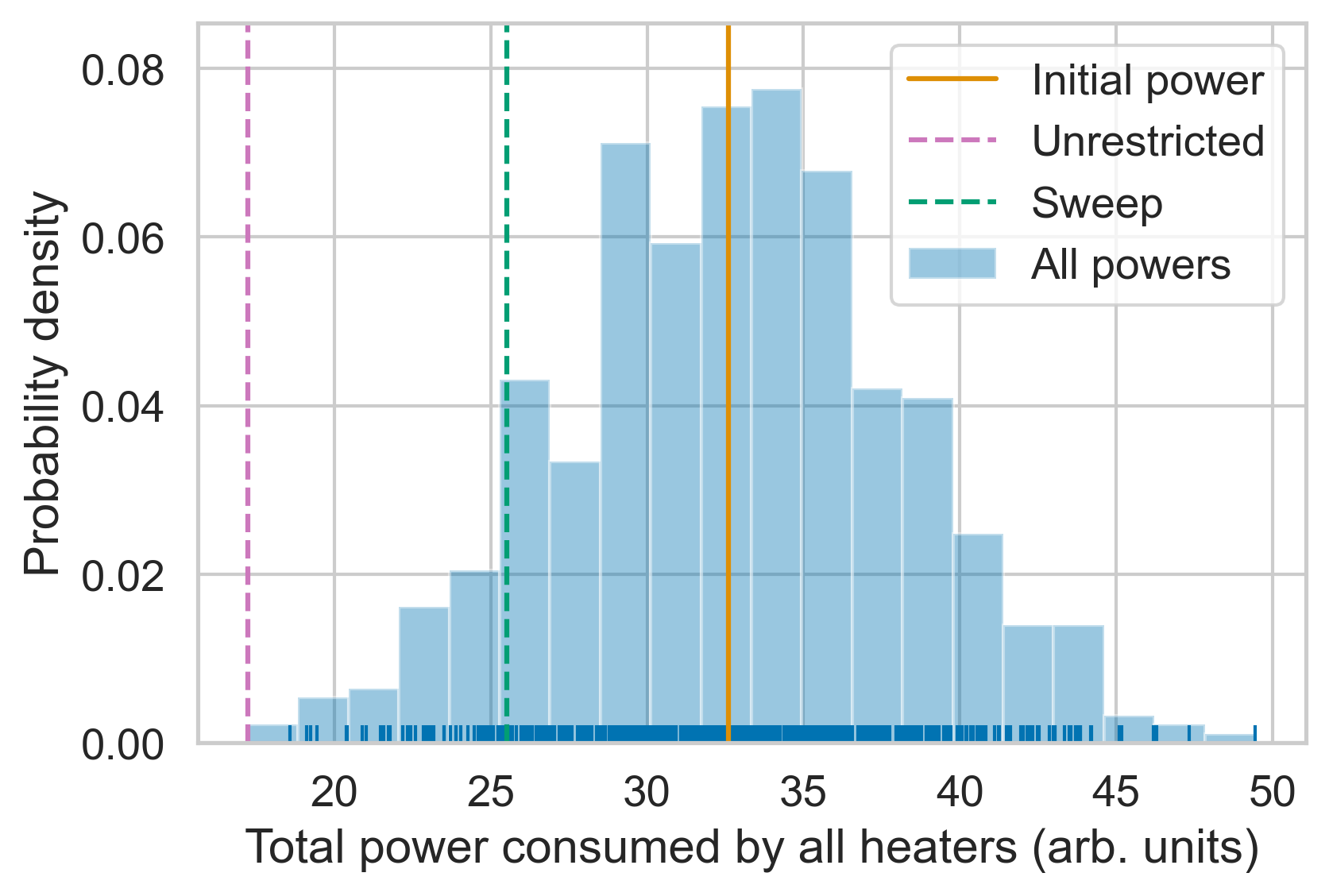}
\caption{\label{Fig:PowerChange} \textbf{Example of change of power via permutations.}
A random 4-mode unitary is drawn from the Haar measure.
Its power consumption on an interferometer is calculated assuming a linear relation between powers and phases with zero powers representing uniform random phases.
The initial power consumption is depicted by the vertical orange line.
The power consumption of the permuted unitary matrices is represented by the short vertical lines at the bottom of the plot.
The histogram depicts the distribution of these powers for the $(4!)^{2} = 576$ different pairs of input and output permutations.
The vertical dashed lines depict lower power consumption values as determined by two different algorithms.}
\end{figure}

The permuted unitary $\widetilde{U}$ will be implemented with different voltages, and it will thus have a different power consumption proportional to $\sum_{\ell} \widetilde{V}_{\ell}^{2}$.
Permuting the input port labels requires permuting the light that is impinged at these ports, which means that the four squeezers will be tuned to perform squeezing $\{s_{2}, s_{3}, s_{1}, s_{4}\}$ respectively.
Likewise, the output statistics need to be permuted as well before being returned to the user.
Suppose that the PNRs return photon counts $\{n^{\prime}_{1}, n^{\prime}_{2}, n^{\prime}_{3}, n^{\prime}_{4}\}$ respectively, then these counts are permuted according to $Q$ and the value returned to the user is $\{n^{\prime}_{1}, n^{\prime}_{2}, n^{\prime}_{4}, n^{\prime}_{3}\}$.
From the standpoint of the user, this complete procedure gives identical outcome probabilities as compared to the original unpermuted procedure, but with a different power consumption.

The different power consumption values of a random unitary matrix across all possible permutations are plotted in \cref{Fig:PowerChange}.	
We note that for random choice of unitary matrix and chip parameters, the power consumption can be decreased or increased around $50\%$ in each direction through permutations. 
The applications to reducing the mean or variance of the power consumption values across all unitary matrices exploits this ability to change the powers of each individual unitary.

\section{Methods for searching over port allocations}
\label{Sec:PPMethods}

In this appendix, we describe in more detail the two algorithms introduced in the main text for searching over the different permutations involved in the port-allocation method.
We note that it is important to consider these methods carefully as the space of possible port allocations scales very fast, as $n!^{2}$ for an $n$-mode unitary.

\subsection{Unrestricted search over all port allocations}

For a small ($n<6$) number of modes of the interferometer, all permutations can be tried and the suitable pairs of input and output port allocations can be returned as determined by the applications mentioned in the main text or future applications.
For instance, for the reduction of power consumption, those allocations that minimize the power consumption are returned.
For reducing the range of power consumption, those allocations that lead to power consumption within the desired range are returned.
Note that this function has a run-time which scales at least as badly as $(n!)^2$, which is the scaling on the number of permutations that need to be tried.

We note that in many cases, we do not need to find the best possible allocation, only one that satisfies the desired criteria.
Hence, we can search randomly for permutations that meet the desired criteria for small-to-medium sized interferometers ($n<20$).
Randomized methods such as the Fisher–Yates shuffle~\cite{Durstenfeld1964} or more recent versions thereof~\cite{Okounkov2000} can be used.
Once a suitable pair of permutations is found, i.e., one with a power consumption below a certain desired threshold (for reducing power consumption), or one with power consumption or fidelity within the desired range (for specified ranges of powers or fidelities), then the search over random permutations can be truncated.
If no such suitable pair of permutations is found within a chosen length of time, then the method can declare failure.
One positive feature of this procedure is that the searching over these random permutations can be performed completely in parallel, which could help speed up this search.

\begin{figure*}
\centering
\subfloat[Trivial permutations]{\label{Fig:TrivialPerms} \includegraphics[width = 0.75\textwidth]{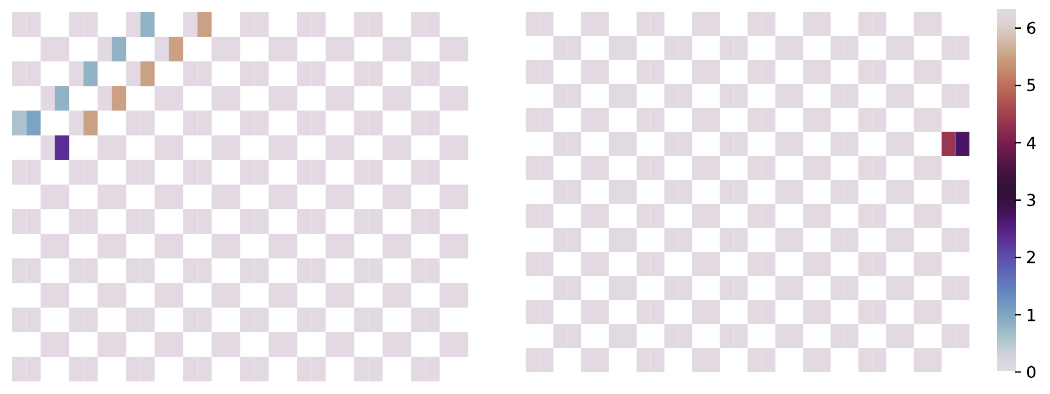}}\\
\subfloat[Non-trivial permutations]{\label{Fig:NonTrivialPerms} \includegraphics[width = 0.75\textwidth]{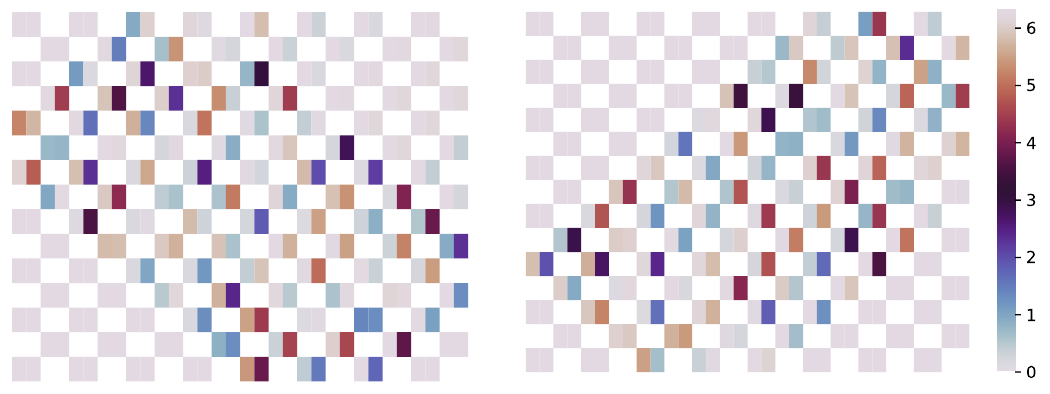}}
\caption{
\textbf{Effect on interferometer phases from trivial and non-trivial permutations.}
We consider a 16 mode interferometer whose input and output ports are numbered according to \cref{Fig:Relabeling}. 
We calculate the internal and external phases of each MZI before and after permutations. 
This change in phases is plotted in these heatmaps.
Input permutations are on the left and output permutations on the right.
(a) \textbf{Trivial permutations}: (left) permute inputs 5 and 6 and (right) permute outputs 6 and 7. 
With these permutations, we see that only a few of the phases change.
(b) \textbf{Non-trivial permutations}: (left) permute inputs 6 and 7 and (right) permute outputs 5 and 6. 
With non-trivial permutations, a large number of phases on the interferometer change.}
\end{figure*}

\subsection{Sweep Search algorithm}

For larger numbers of modes (say between \num{10} and \num{100}), a different searching algorithm can be used.
This algorithm relies on identifying that some permutations have a bigger impact on the interferometer phases than others.

Consider nearest neighbour transpositions, which are permutations that swap two neighbouring modes, say $i$ and $i+1$.
These permutations can be classified as either ``trivial'' or ``non-trivial'' permutations based on how they affect the phases on the interferometer. 

Trivial permutations are those that affect only a few interferometer phases. 
As an example, consider the action of exchanging neighbouring input ports  $i$ and $i+1$ of an interferometer.
If the first layer of beam-splitters in the chip includes a beam-splitter that acts on these two modes, then the permutation will only change the internal and external phase of this single beam-splitter phase and a few other external phases and the rest of the interferometer will be left unchanged.
After using the Clements \textit{et al.}~decomposition~\cite{clements2016optimal}, all the input transpositions swapping modes $i$ and $i+1$ with $i$ odd are trivial, and so are their compositions.
The same applies for output port transpositions if the total number of modes is odd, but if the number of modes is even, then the output port transpositions are trivial for $i$ even.
This is illustrated in \cref{Fig:TrivialPerms} where swapping input modes 5 and 6 and output modes 6 and 7 change only a few phases.

The remaining nearest neighbour transpositions are all non-trivial in their action, i.e., transpositions that relabel ports connected to different beam-splitters have a non-trivial action on all the phases. 
This is also illustrated in \cref{Fig:NonTrivialPerms}, where swapping input modes 6 and 7 and output modes 5 and 6 change a large portion of the interferometer phases.
The same holds for all compositions of the non-trivial transpositions.

Two different trivial permutations act on nearest neighbour modes without acting on common elements, they commute amongst themselves, so they can be applied in any order.
Similarly, the non-trivial permutations can be applied in any order.
Note that trivial and purely non-trivial permutations so defined do not cover the entire set of permutations.
In other words, there are permutations, which are neither trivial nor purely non-trivial but actually comprise a mix of trivial and non-trivial transpositions.
That said, any permutation can be obtained by acting purely trivial and purely non-trivial transpositions one after the other.

In order to speed up the search over permutations, the Sweep Search algorithm chooses to focus the search in the space of non-trivial permutations. 
It first searches over all purely non-trivial permutations of a given unitary for the one that minimizes power consumption.
Once such a permutation is found, then this permuted unitary is the starting point of the next step, in which all trivial permutations are searched over to find the one that minimizes the power consumption.
Multiple such repetitions can be performed in order to obtain a better permutation.
The repetitions can be stopped once the constraints are met either in terms of a low enough power or distance from target unitary being achieved, or a power in the desired range being obtained, or finally if a maximum allowed computation time is exceeded.

This heuristic algorithm scales at least as fast as $k 2^n$, where $k$ is the number of repetitions performed and $2^{n/2}$ is the number of trivial or purely non-trivial permutations acting exclusively on input ports or output ports.
In practice a value of $k = 2$ or $3$ performs as well as higher values.
Thus, for a constant value of $k$, the scaling of this algorithm is significantly faster than the scaling of the unrestricted search.
Indeed, this faster scaling allows us to search over permutations of much larger unitary matrices.

\begin{table*}
\caption{Comparison of Clements~\textit{et al.} decomposition and tailored compilation (this work)}
\begin{tabular}{p{2.5cm} | l | l}
\toprule
& Clements~\textit{et al.}~\cite{clements2016optimal} & Tailored compilation  \\
\midrule
Nulling matrices  &
$\begin{aligned}[c]
i \e^{i \frac{\phi_{\text{i}}}{2}}
\begin{pmatrix*}[c]
\e^{i \phi_{\text{e}}} \, \sin{\frac{\phi_{\text{i}}}{2}} & \cos{\frac{\phi_{\text{i}}}{2}} \\
\e^{i \phi_{\text{e}}} \, \cos{\frac{\phi_{\text{i}}}{2}} & - \sin{\frac{\phi_{\text{i}}}{2}}
\end{pmatrix*}
\end{aligned}$ &
$\begin{aligned}[c]
 \begin{pmatrix*}[c]
 \e^{i \phi_{\text{e}}} \left(\e^{i \phi_{\text{i}}} \cos^{2}{\theta} - \sin^{2}{\theta}\right)  & i \e^{i \frac{\phi_{\text{i}}}{2} } \cos \frac{\phi_{\text{i}}}{2}  \sin{2 \theta}\\
  i \e^{i \left( \phi_{\text{e}} + \frac{\phi_{\text{i}}}{2} \right)}  \cos \frac{\phi_{\text{i}}}{2}  \sin{2 \theta} & - \e^{i \phi_{\text{i}}} \sin^{2}{\theta} + \cos^{2}{\theta}
 \end{pmatrix*}
\end{aligned}$
\\
\midrule
\vspace*{4mm}&
&
\(\displaystyle s := \frac{1}{\sin2 \theta} \cos \left( \tan^{-1} \left| \frac{ U_{n-1, m}} {U_{n, m}} \right| \right) \),
\\
Nulling equations (with $Z$)\vspace*{2.5mm}
&
\(\displaystyle \phi_{\text{i}} = 2 \tan^{-1} \left| \frac{ U_{n-1, m}} {U_{n, m}} \right| \)
&
\(\displaystyle \phi_{\text{i}} =
\begin{cases}
	2 \cos^{-1}s & \text{if $-1 < s < 1$}\\
	2 \pi &  \text{if $s < -1$} \\
	0 &  \text{if $s > 1$} \\
\end{cases} \)
\\
&
\(\displaystyle \phi_{\text{e}} = - \arg \left( \frac{ U_{n-1, m}} {U_{n, m}} \right) \)
&
\(\displaystyle \phi_{\text{e}} = - \arg \left( \frac{ U_{n-1, m}} {U_{n, m}} \right) + \frac{\pi}{2} - \frac{\phi_{\text{i}}}{2} + \tan^{-1} \left( \frac{-\sin^{2} \theta \sin \phi_{\text{i}}} {\cos^{2} \theta - \sin^{2} \theta  \cos \phi_{\text{i}}} \right) \)
\\
\midrule
\vspace*{4mm}
&
&
\(\displaystyle s := \frac{1}{\sin 2\theta} \cos \left( \tan^{-1} \left| \frac{ U_{n, m+1}} {U_{n, m}} \right| \right) \),
\\
Nulling equations (with $Z^{-1}$)\vspace*{3mm}
&
\(\displaystyle \phi_{\text{i}} = 2 \tan^{-1} \left| \frac{ U_{n, m+1}} {U_{n, m}} \right| \)
&
\(\displaystyle \phi_{\text{i}} =
\begin{cases}
	2 \cos^{-1}s & \text{if $-1 < s < 1$}\\
	2 \pi &  \text{if $s < -1$} \\
	0 &  \text{if $s > 1$} \\
\end{cases} \)
\\
&
\(\displaystyle \phi_{\text{e}} = - \arg \left( \frac{ U_{n, m+1}} {U_{n, m}} \right) + \pi \)
&
\(\displaystyle \phi_{\text{e}} = - \arg \left( \frac{ U_{n, m+1}} {U_{n, m}}\right) - \frac{\pi}{2} + \frac{\phi_{\text{i}}}{2} + \tan^{-1} \left( \frac{-\cos^{2} \theta \sin \phi_{\text{i}}} {-\sin^{2} \theta + \cos^{2} \theta  \cos \phi_{\text{i}}} \right) \)
\\
\bottomrule
\end{tabular}
\label{Tab:Comparision}
\end{table*}

\section{Details tailored compilation}
\label{Sec:Trusplit}

Here, we consider a provide more details about our tailored compilation procedure.
As described in the main text, this procedure accounts for imbalanced beam-splitters by decomposing the $N \times N$ unitary $U$ using matrices corresponding to $2 \times 2$ imperfect MZIs rather than those corresponding to perfect MZIs.
To realistically model these imperfect interferometers, we consider the effective transformation implemented by the imperfect MZIs.
Non-balanced beam-splitters can be modelled as
\begin{align}
B(\theta) =
	\begin{pmatrix*}[c]
	\cos{\theta} & i \sin{\theta} \\
	i \sin{\theta} & \cos{\theta}
	\end{pmatrix*}
	\label{Eq:BS}
\end{align}
whose reflectivity is $\cos^{2} \theta$.
This reduces to \cref{Eq:B_bal} when $\theta = \pi/4$.

The phase-shifters are the same, as in \cref{Eq:Phase}.
Multiplying them together, we find the MZ unit $Z^{\theta}$ that is actually implemented when the beam-splitters are imperfect, i.e.,
\begin{align}
Z^{\theta} &= B(\theta) R(\phi_{\text{i}}) B(\theta) R(\phi_{\text{e}}) \\
&=
\begin{pmatrix*}[c]
\e^{i \phi_{\text{e}}} \left(\e^{i \phi_{\text{i}}} \cos^{2}{\theta} - \sin^{2}{\theta}\right)  & i \e^{i \frac{\phi_{\text{i}}}{2} } \cos\frac{\phi_{\text{i}}}{2} \sin{2 \theta}\\
 i \e^{i \left( \phi_{\text{e}} + \frac{\phi_{\text{i}}}{2} \right)}  \cos \frac{\phi_{\text{i}}}{2} \sin{2 \theta} & - \e^{i \phi_{\text{i}}} \sin^{2}{\theta} + \cos^{2}{\theta}
\end{pmatrix*}.
\label{Eq:MZ_Trusplit}
\end{align}
Hence, a decomposition following the nulling order of the Clements~\emph{et al.} procedure is performed to account for the imperfections in the decomposition procedure itself.
The comparison of the equations involved in the decomposition is provided in \cref{Tab:Comparision}.

\begin{figure}
\includegraphics[width=\columnwidth]{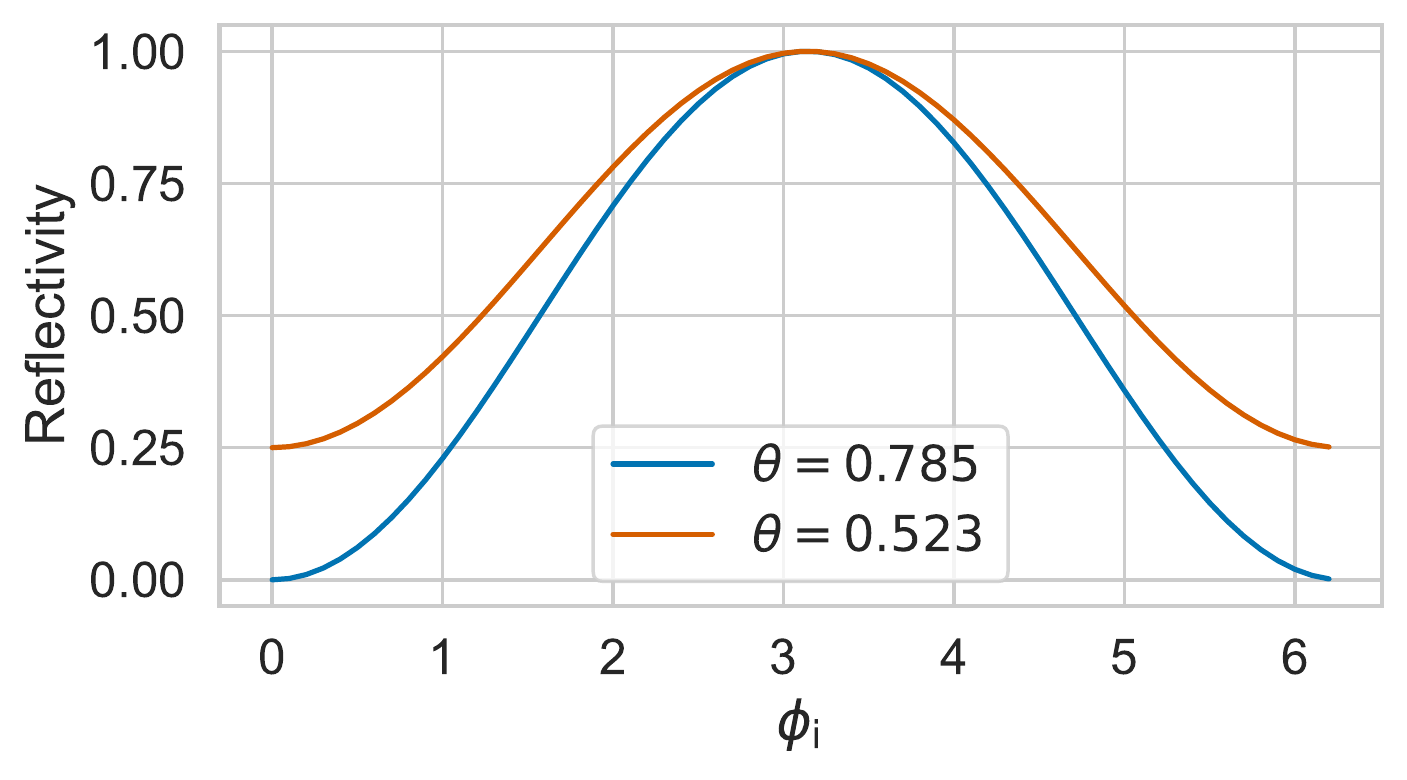}
\caption{\label{Fig:RefRange}
Range of reflectivities that can be implemented by an MZI by varying the internal angle $\phi_{\text{i}}$ for different values of splitting values.
When $\theta = \pi/4$, all reflectivity values ranging from $0$ to $1$ can be implemented, but when $\theta$ differs from the ideal value of $\pi/4$, only a smaller range of reflectivities, e.g., from $0.25$ to $1$ can be implemented.}
\end{figure}

We also note that despite using the tailored decomposition, there are certain operators that cannot be implemented on the imperfect chips.
For instance, for a splitting ratio characterized by $\theta$, it is not possible to implement transformations that require a reflectivity of less than $\cos^{2} 2\theta$, as can be seen from \cref{Fig:RefRange}.
In such situations, the fidelity of a unitary operator can be improved with port allocation. 
Port allocation can lead to permutations such that MZIs requiring low values of reflectivity can be replaced by MZIs requiring larger values of reflectivities, which can be implemented exactly.

\section{Details on calibration procedures to obtain splitting ratios}
\label{Sec:Calibration}

In this section, we provide more details about the procedures used in the calibration of the splitting ratios.
In particular, these procedures assume access to the chip with un-entangled states of classical or quantum light at the inputs and intensity measurements at the outputs.

\subsection{Method 1: Finding splitting ratio of each MZI}
The first method to find the splitting ratios of each MZI is based on the observation that in the presence of imperfect splitting ratios, the ``bar'' state of an MZI (i.e., all light going straight through) can be implemented perfectly, whereas the ``cross'' state (i.e., all light from one port completely crossing over to the other port) cannot.
From \cref{Eq:MZ_Trusplit}, we infer that the bar state is achieved when $\phi_{\text{i}} = \pi$ and the cross state is achieved when $\phi_{\text{i}} = 0$.

The procedure to find the splitting ratio of a particular MZI is as follows:
\begin{enumerate}
\item
The splitting ratio of a particular MZI can be found by setting that MZI to the cross state and every other MZI to the bar state.
\item 
Light is incident to one of the input ports of the crossed MZI and the intensity at both its output ports is measured.
\item 
Using \cref{Eq:MZ_Trusplit}, the intensity at the two output ports is proportional to $\cos^{2} 2 \theta$ and $\sin^{2} 2 \theta$.
By measuring the intensities and calculating their ratio $\cot^{2} 2 \theta$, infer the value of $\theta$ for that MZI.
\item
Repeat the above steps for each MZI to find all the splitting ratios.
\end{enumerate}

However, this method might not yield the splitting ratios to a high precision in the presence of other imperfections such as imperfect calibration of the voltage-phase relationship, because in this case there might be imprecision in the programmed phase value.
In particular, the method relies on the possibility of implementing the bar state perfectly in each MZI in the interferometer, which might not hold perfectly because of defects other than those that we consider here.
So we devise another method to find the mean of the splitting ratio distribution to high precision using optimization.

\subsection{Method 2: Finding best splitting ratio of all MZIs together}

The second method for the calibration of the splitting ratios relies on using the tailored compilation itself and optimizing to find that value of splitting ratio which leads to the best performance of the compiler.
The method proceeds as follows:
\begin{enumerate}
\item
First guess an initial splitting ratio and using this, implement a specific unitary operator $U_{0}$ on the interferometer.
\item
Light is incident on one input port and the intensity at all output ports is measured. This step is repeated for all input ports.
\item
Optimization methods are used to minimize the norm of the difference between the measured and expected intensities while iterating over different values of splitting ratios.
\end{enumerate}
Depending on the size of the interferometer, the optimization can be performed to find one or multiple splitting ratios to best describe the chip imperfections.

For best optimization results, the choice of the specific unitary $U_{0}$ is made such that the gradient in the difference of observed and measured intensities is steepest.
Such a matrix is one in which all the MZIs are set to a value close to the ``cross'' state. 
The motivation for this choice comes from considering \cref{Eq:MZ_Trusplit} from which we see that the norm of the vector of intensity differences between an imperfect MZI with angle $\theta$ and a perfect 50-50 MZI to be equal to $ \frac{1}{8} \left(\cos{\phi_{\text{i}}} + 1\right)^{2} \left(\cos{4 \theta} + 1\right)^{2}$.
We see that this is maximum when $\phi_{\text{i}} = 0$, or the MZI is set to ``cross''.
However, a perfectly ``crossed'' MZI is not possible to implement when the beam-splitters have imperfect splitting ratios.
So, the unitary matrix with the largest difference in actual and expected intensities depends on the actual splitting ratio.
A matrix close to such a matrix can be chosen if a crude initial estimate of the splitting ratios is available. 
In the absence of such an estimate, a small reflectivity value such as 0.2 is observed to perform well (see, for example, \cref{Fig:Pipeline,Fig:PipelineErrors}).

\begin{figure}
\includegraphics[width=\columnwidth]{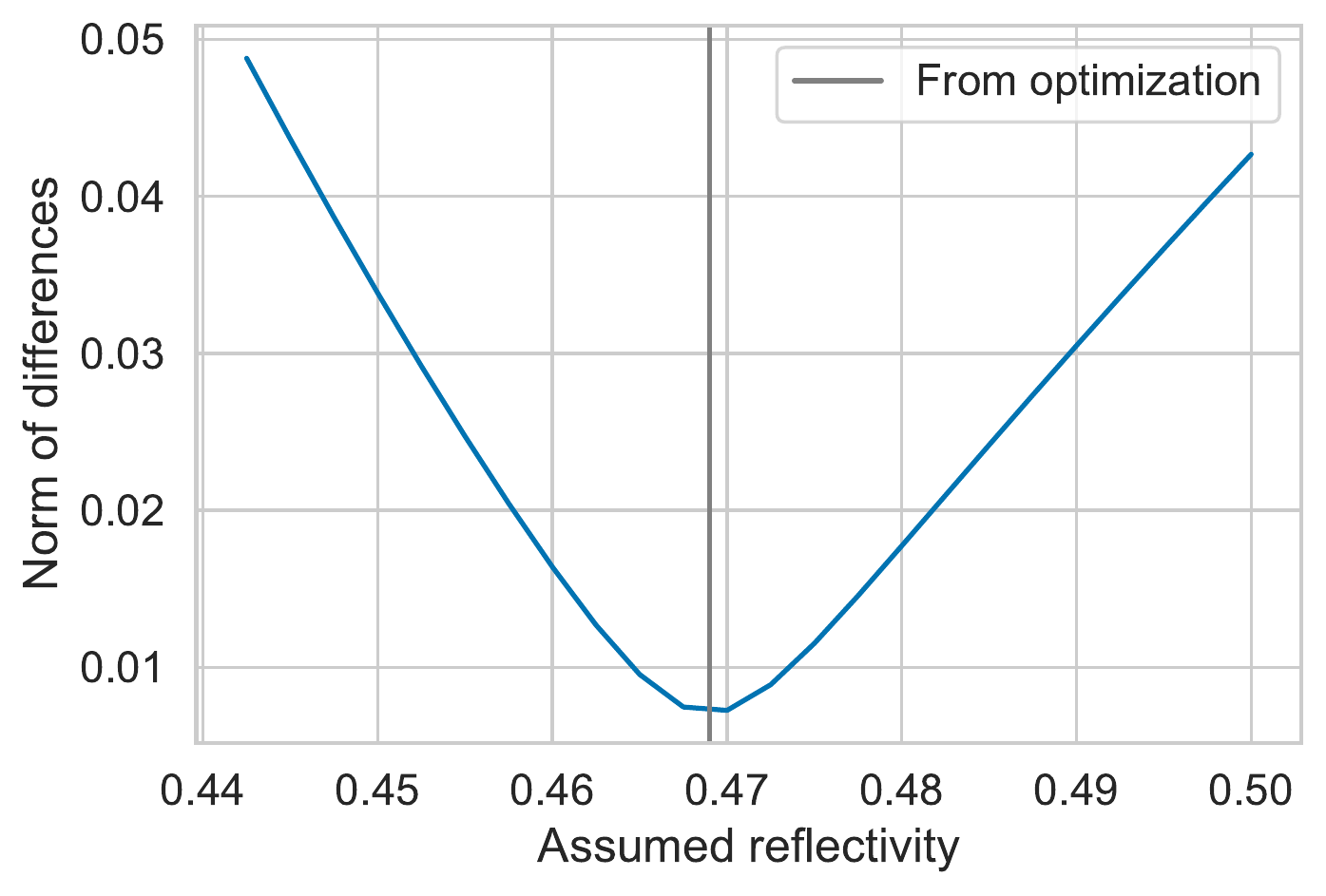}
\caption{The norm of the differences between the expected and observed intensities as a function of the reflectivity assumed by the tailored decomposition.
We consider a 12-mode interferometer, in which the MZI splitting ratios were drawn from a distribution with mean $0.47$ and std dev $0.005$;  light is incident at the first mode. 
As expected, the tailored decomposition reflectivity that minimizes the distance is found to be at $0.469$ using Brent optimization by evaluating the function at only 9 points.}
\label{Fig:Calibration}
\end{figure}

\section{Robustness to non-uniformity in splitting ratios}
\label{Sec:NonUniform}

Here we analyse the robustness of the procedures presented in the main text to the non-uniformity of the splitting ratios of the beam-splitters.
The results presented in the main text focus on the case of each beam-splitter having the same splitting ratio say $47$:$53$, which could be different from $50$:$50$.
However, in reality, there could be some spread of the splitting ratios about this imperfect value because of further fabrication defects.

First, we show that there is a minimum of the fitness function that we are using (i.e., the norm of the difference between the expected and observed intensities) and that numerical methods can find this minimum straightforwardly.
More importantly, this is true even when the actual splitting ratios are not the same across the chip. 
In \cref{Fig:Calibration}, we calculate this norm for different assumed reflectivites and confirm that when the splitting ratios across the interferometer are normally distributed, we do indeed find a minimum in this norm that is close to the mean of the normal distribution.
Using Brent optimization, we find the value of this minimum to be $0.469$, which is close to the actual minimum of $0.47$~\cite{brent1971algorithm}.
Such a gradient descent algorithm needs to perform the decomposition only a few times to arrive at the position of the minimum.

\begin{figure}
\includegraphics[width=\columnwidth]{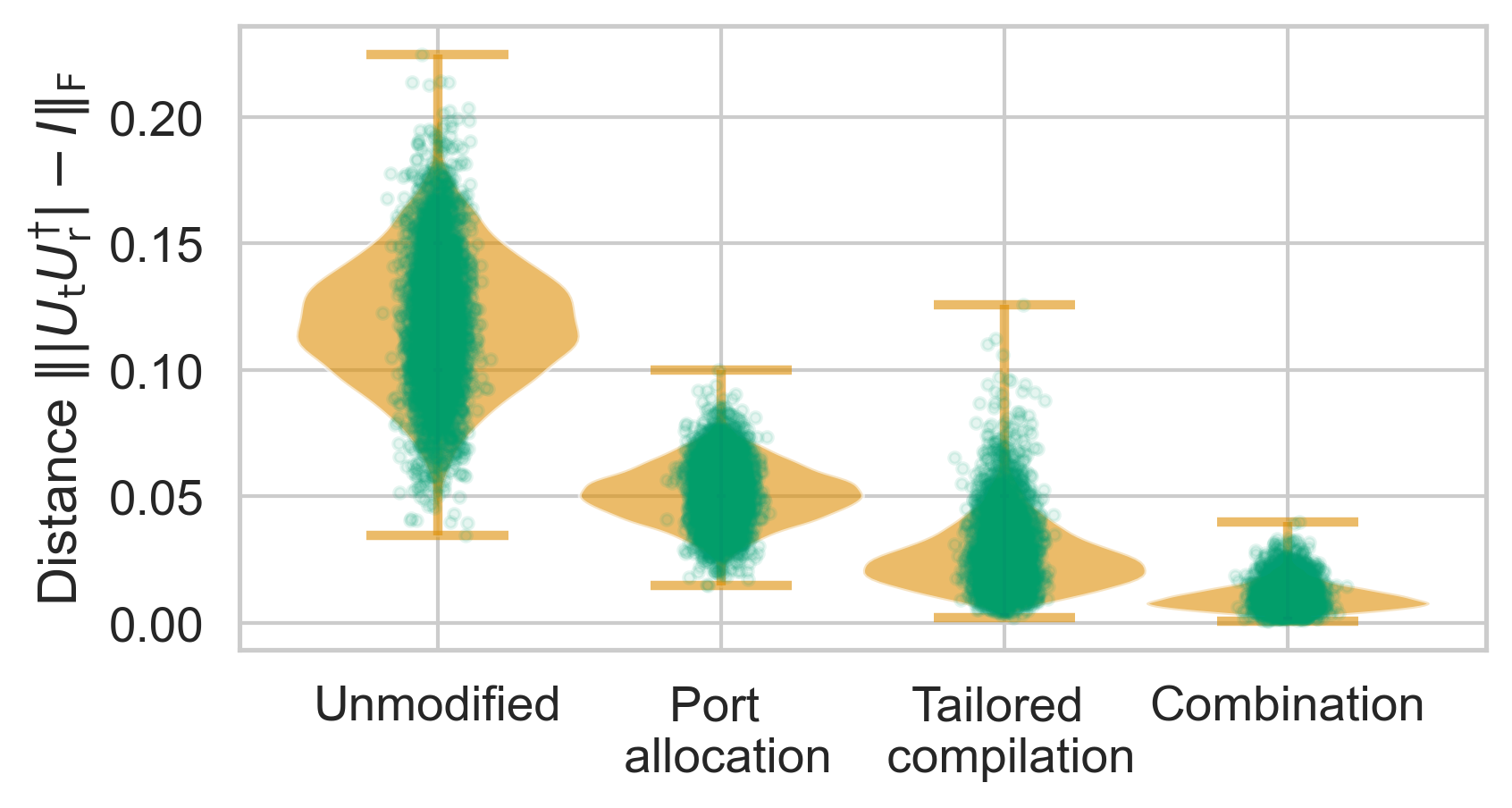}
\caption{\label{Fig:PipelineErrors}Complete work-flow to mitigate imperfect splitting ratios when splitting ratios are all different and unknown.
The beam-splitter reflectivity values of a 4-mode chip are normally distributed with mean $0.47$ and SD $0.005$ but is unknown to a user. \num{5000} unitaries drawn from the Haar measure are implemented on a random chip and a single value to estimate the splitting ratio is determined using the procedure described, and port allocation and tailored compilation are applied individually and in combination to improve the accuracy of the implemented transformation.
}
\end{figure}

To check the performance of the different methods under non-uniform splitting ratios, we perform the simulations of \cref{Fig:Pipeline} but this time assuming a spread in the splitting ratios around the mean value. 
Each of the procedures, namely calibration, port-allocation and tailored compilation are performed with the incorrect assumption that the splitting ratios are all the same.
The results for this simulation are plotted in \cref{Fig:PipelineErrors}.
We observe that despite the non-uniformity in the splitting ratios (and without modifying the procedures to account for this non-uniformity), the methods perform well.
In particular, the combination of the tailored decomposition and the port allocation provides a factor $11.5$ enhancement in the distance of the implemented unitary transformations.

\begin{thebibliography}{48}%
\makeatletter
\providecommand \@ifxundefined [1]{%
 \@ifx{#1\undefined}
}%
\providecommand \@ifnum [1]{%
 \ifnum #1\expandafter \@firstoftwo
 \else \expandafter \@secondoftwo
 \fi
}%
\providecommand \@ifx [1]{%
 \ifx #1\expandafter \@firstoftwo
 \else \expandafter \@secondoftwo
 \fi
}%
\providecommand \natexlab [1]{#1}%
\providecommand \enquote  [1]{``#1''}%
\providecommand \bibnamefont  [1]{#1}%
\providecommand \bibfnamefont [1]{#1}%
\providecommand \citenamefont [1]{#1}%
\providecommand \href@noop [0]{\@secondoftwo}%
\providecommand \href [0]{\begingroup \@sanitize@url \@href}%
\providecommand \@href[1]{\@@startlink{#1}\@@href}%
\providecommand \@@href[1]{\endgroup#1\@@endlink}%
\providecommand \@sanitize@url [0]{\catcode `\\12\catcode `\$12\catcode
  `\&12\catcode `\#12\catcode `\^12\catcode `\_12\catcode `\%12\relax}%
\providecommand \@@startlink[1]{}%
\providecommand \@@endlink[0]{}%
\providecommand \url  [0]{\begingroup\@sanitize@url \@url }%
\providecommand \@url [1]{\endgroup\@href {#1}{\urlprefix }}%
\providecommand \urlprefix  [0]{URL }%
\providecommand \Eprint [0]{\href }%
\providecommand \doibase [0]{https://doi.org/}%
\providecommand \selectlanguage [0]{\@gobble}%
\providecommand \bibinfo  [0]{\@secondoftwo}%
\providecommand \bibfield  [0]{\@secondoftwo}%
\providecommand \translation [1]{[#1]}%
\providecommand \BibitemOpen [0]{}%
\providecommand \bibitemStop [0]{}%
\providecommand \bibitemNoStop [0]{.\EOS\space}%
\providecommand \EOS [0]{\spacefactor3000\relax}%
\providecommand \BibitemShut  [1]{\csname bibitem#1\endcsname}%
\let\auto@bib@innerbib\@empty
\bibitem [{\citenamefont {Bourassa}\ \emph {et~al.}(2021)\citenamefont
  {Bourassa}, \citenamefont {Alexander}, \citenamefont {Vasmer}, \citenamefont
  {Patil}, \citenamefont {Tzitrin}, \citenamefont {Matsuura}, \citenamefont
  {Su}, \citenamefont {Baragiola}, \citenamefont {Guha}, \citenamefont
  {Dauphinais}, \citenamefont {Sabapathy}, \citenamefont {Menicucci},\ and\
  \citenamefont {Dhand}}]{bourassa2021blueprint}%
  \BibitemOpen
  \bibfield  {author} {\bibinfo {author} {\bibfnamefont {J.~E.}\ \bibnamefont
  {Bourassa}}, \bibinfo {author} {\bibfnamefont {R.~N.}\ \bibnamefont
  {Alexander}}, \bibinfo {author} {\bibfnamefont {M.}~\bibnamefont {Vasmer}},
  \bibinfo {author} {\bibfnamefont {A.}~\bibnamefont {Patil}}, \bibinfo
  {author} {\bibfnamefont {I.}~\bibnamefont {Tzitrin}}, \bibinfo {author}
  {\bibfnamefont {T.}~\bibnamefont {Matsuura}}, \bibinfo {author}
  {\bibfnamefont {D.}~\bibnamefont {Su}}, \bibinfo {author} {\bibfnamefont
  {B.~Q.}\ \bibnamefont {Baragiola}}, \bibinfo {author} {\bibfnamefont
  {S.}~\bibnamefont {Guha}}, \bibinfo {author} {\bibfnamefont {G.}~\bibnamefont
  {Dauphinais}}, \bibinfo {author} {\bibfnamefont {K.~K.}\ \bibnamefont
  {Sabapathy}}, \bibinfo {author} {\bibfnamefont {N.~C.}\ \bibnamefont
  {Menicucci}},\ and\ \bibinfo {author} {\bibfnamefont {I.}~\bibnamefont
  {Dhand}},\ }\bibfield  {title} {\bibinfo {title} {Blueprint for a {S}calable
  {P}hotonic {F}ault-{T}olerant {Q}uantum {C}omputer},\ }\href
  {https://doi.org/10.22331/q-2021-02-04-392} {\bibfield  {journal} {\bibinfo
  {journal} {{Quantum}}\ }\textbf {\bibinfo {volume} {5}},\ \bibinfo {pages}
  {392} (\bibinfo {year} {2021})}\BibitemShut {NoStop}%
\bibitem [{\citenamefont {Rudolph}(2017)}]{Rudolph2017a}%
  \BibitemOpen
  \bibfield  {author} {\bibinfo {author} {\bibfnamefont {T.}~\bibnamefont
  {Rudolph}},\ }\bibfield  {title} {\bibinfo {title} {{Why I am optimistic
  about the silicon-photonic route to quantum computing}},\ }\href
  {https://doi.org/10.1063/1.4976737} {\bibfield  {journal} {\bibinfo
  {journal} {APL Photonics}\ }\textbf {\bibinfo {volume} {2}},\ \bibinfo
  {pages} {030901} (\bibinfo {year} {2017})}\BibitemShut {NoStop}%
\bibitem [{\citenamefont {Aaronson}\ and\ \citenamefont
  {Arkhipov}(2013)}]{Aaronson2013a}%
  \BibitemOpen
  \bibfield  {author} {\bibinfo {author} {\bibfnamefont {S.}~\bibnamefont
  {Aaronson}}\ and\ \bibinfo {author} {\bibfnamefont {A.}~\bibnamefont
  {Arkhipov}},\ }\bibfield  {title} {\bibinfo {title} {{The Computational
  Complexity of Linear Optics}},\ }\href
  {https://doi.org/10.4086/toc.2013.v009a004} {\bibfield  {journal} {\bibinfo
  {journal} {Theory Comput.}\ }\textbf {\bibinfo {volume} {9}},\ \bibinfo
  {pages} {143} (\bibinfo {year} {2013})}\BibitemShut {NoStop}%
\bibitem [{\citenamefont {Zhong}\ \emph {et~al.}(2020)\citenamefont {Zhong},
  \citenamefont {Wang}, \citenamefont {Deng}, \citenamefont {Chen},
  \citenamefont {Peng}, \citenamefont {Luo}, \citenamefont {Qin}, \citenamefont
  {Wu}, \citenamefont {Ding}, \citenamefont {Hu} \emph {et~al.}}]{Zhong2020a}%
  \BibitemOpen
  \bibfield  {author} {\bibinfo {author} {\bibfnamefont {H.-S.}\ \bibnamefont
  {Zhong}}, \bibinfo {author} {\bibfnamefont {H.}~\bibnamefont {Wang}},
  \bibinfo {author} {\bibfnamefont {Y.-H.}\ \bibnamefont {Deng}}, \bibinfo
  {author} {\bibfnamefont {M.-C.}\ \bibnamefont {Chen}}, \bibinfo {author}
  {\bibfnamefont {L.-C.}\ \bibnamefont {Peng}}, \bibinfo {author}
  {\bibfnamefont {Y.-H.}\ \bibnamefont {Luo}}, \bibinfo {author} {\bibfnamefont
  {J.}~\bibnamefont {Qin}}, \bibinfo {author} {\bibfnamefont {D.}~\bibnamefont
  {Wu}}, \bibinfo {author} {\bibfnamefont {X.}~\bibnamefont {Ding}}, \bibinfo
  {author} {\bibfnamefont {Y.}~\bibnamefont {Hu}}, \emph {et~al.},\ }\bibfield
  {title} {\bibinfo {title} {Quantum computational advantage using photons},\
  }\href {https://doi.org/10.1126/science.abe8770} {\bibfield  {journal}
  {\bibinfo  {journal} {Science}\ }\textbf {\bibinfo {volume} {370}},\ \bibinfo
  {pages} {1460} (\bibinfo {year} {2020})}\BibitemShut {NoStop}%
\bibitem [{\citenamefont {Hamilton}\ \emph {et~al.}(2017)\citenamefont
  {Hamilton}, \citenamefont {Kruse}, \citenamefont {Sansoni}, \citenamefont
  {Barkhofen}, \citenamefont {Silberhorn},\ and\ \citenamefont
  {Jex}}]{Hamilton2017}%
  \BibitemOpen
  \bibfield  {author} {\bibinfo {author} {\bibfnamefont {C.~S.}\ \bibnamefont
  {Hamilton}}, \bibinfo {author} {\bibfnamefont {R.}~\bibnamefont {Kruse}},
  \bibinfo {author} {\bibfnamefont {L.}~\bibnamefont {Sansoni}}, \bibinfo
  {author} {\bibfnamefont {S.}~\bibnamefont {Barkhofen}}, \bibinfo {author}
  {\bibfnamefont {C.}~\bibnamefont {Silberhorn}},\ and\ \bibinfo {author}
  {\bibfnamefont {I.}~\bibnamefont {Jex}},\ }\bibfield  {title} {\bibinfo
  {title} {Gaussian boson sampling},\ }\href
  {https://doi.org/10.1103/PhysRevLett.119.170501} {\bibfield  {journal}
  {\bibinfo  {journal} {Phys. Rev. Lett.}\ }\textbf {\bibinfo {volume} {119}},\
  \bibinfo {pages} {170501} (\bibinfo {year} {2017})},\ \Eprint
  {https://arxiv.org/abs/1612.01199} {arXiv:1612.01199} \BibitemShut {NoStop}%
\bibitem [{\citenamefont {Kruse}\ \emph {et~al.}(2019)\citenamefont {Kruse},
  \citenamefont {Hamilton}, \citenamefont {Sansoni}, \citenamefont {Barkhofen},
  \citenamefont {Silberhorn},\ and\ \citenamefont {Jex}}]{Kruse2019}%
  \BibitemOpen
  \bibfield  {author} {\bibinfo {author} {\bibfnamefont {R.}~\bibnamefont
  {Kruse}}, \bibinfo {author} {\bibfnamefont {C.~S.}\ \bibnamefont {Hamilton}},
  \bibinfo {author} {\bibfnamefont {L.}~\bibnamefont {Sansoni}}, \bibinfo
  {author} {\bibfnamefont {S.}~\bibnamefont {Barkhofen}}, \bibinfo {author}
  {\bibfnamefont {C.}~\bibnamefont {Silberhorn}},\ and\ \bibinfo {author}
  {\bibfnamefont {I.}~\bibnamefont {Jex}},\ }\bibfield  {title} {\bibinfo
  {title} {A detailed study of {{Gaussian Boson Sampling}}},\ }\href
  {https://doi.org/10.1103/PhysRevA.100.032326} {\bibfield  {journal} {\bibinfo
   {journal} {Phys. Rev. A}\ }\textbf {\bibinfo {volume} {100}},\ \bibinfo
  {pages} {032326} (\bibinfo {year} {2019})}\BibitemShut {NoStop}%
\bibitem [{\citenamefont {Deshpande}\ \emph {et~al.}(2021)\citenamefont
  {Deshpande}, \citenamefont {Mehta}, \citenamefont {Vincent}, \citenamefont
  {Quesada}, \citenamefont {Hinsche}, \citenamefont {Ioannou}, \citenamefont
  {Madsen}, \citenamefont {Lavoie}, \citenamefont {Qi}, \citenamefont {Eisert},
  \citenamefont {Hangleiter}, \citenamefont {Fefferman},\ and\ \citenamefont
  {Dhand}}]{deshpande2021quantum}%
  \BibitemOpen
  \bibfield  {author} {\bibinfo {author} {\bibfnamefont {A.}~\bibnamefont
  {Deshpande}}, \bibinfo {author} {\bibfnamefont {A.}~\bibnamefont {Mehta}},
  \bibinfo {author} {\bibfnamefont {T.}~\bibnamefont {Vincent}}, \bibinfo
  {author} {\bibfnamefont {N.}~\bibnamefont {Quesada}}, \bibinfo {author}
  {\bibfnamefont {M.}~\bibnamefont {Hinsche}}, \bibinfo {author} {\bibfnamefont
  {M.}~\bibnamefont {Ioannou}}, \bibinfo {author} {\bibfnamefont
  {L.}~\bibnamefont {Madsen}}, \bibinfo {author} {\bibfnamefont
  {J.}~\bibnamefont {Lavoie}}, \bibinfo {author} {\bibfnamefont
  {H.}~\bibnamefont {Qi}}, \bibinfo {author} {\bibfnamefont {J.}~\bibnamefont
  {Eisert}}, \bibinfo {author} {\bibfnamefont {D.}~\bibnamefont {Hangleiter}},
  \bibinfo {author} {\bibfnamefont {B.}~\bibnamefont {Fefferman}},\ and\
  \bibinfo {author} {\bibfnamefont {I.}~\bibnamefont {Dhand}},\ }\href@noop {}
  {\bibinfo {title} {Quantum computational supremacy via high-dimensional
  gaussian boson sampling}} (\bibinfo {year} {2021}),\ \Eprint
  {https://arxiv.org/abs/2102.12474} {arXiv:2102.12474 [quant-ph]} \BibitemShut
  {NoStop}%
\bibitem [{\citenamefont {Duan}\ \emph {et~al.}(2001)\citenamefont {Duan},
  \citenamefont {Lukin}, \citenamefont {Cirac},\ and\ \citenamefont
  {Zoller}}]{Duan2001a}%
  \BibitemOpen
  \bibfield  {author} {\bibinfo {author} {\bibfnamefont {L.-M.}\ \bibnamefont
  {Duan}}, \bibinfo {author} {\bibfnamefont {M.}~\bibnamefont {Lukin}},
  \bibinfo {author} {\bibfnamefont {J.~I.}\ \bibnamefont {Cirac}},\ and\
  \bibinfo {author} {\bibfnamefont {P.}~\bibnamefont {Zoller}},\ }\bibfield
  {title} {\bibinfo {title} {Long-distance quantum communication with atomic
  ensembles and linear optics},\ }\href {https://doi.org/10.1038/35106500}
  {\bibfield  {journal} {\bibinfo  {journal} {Nature}\ }\textbf {\bibinfo
  {volume} {414}},\ \bibinfo {pages} {413} (\bibinfo {year}
  {2001})}\BibitemShut {NoStop}%
\bibitem [{\citenamefont {Motes}\ \emph {et~al.}(2015)\citenamefont {Motes},
  \citenamefont {Olson}, \citenamefont {Rabeaux}, \citenamefont {Dowling},
  \citenamefont {Olson},\ and\ \citenamefont {Rohde}}]{Motes2015a}%
  \BibitemOpen
  \bibfield  {author} {\bibinfo {author} {\bibfnamefont {K.~R.}\ \bibnamefont
  {Motes}}, \bibinfo {author} {\bibfnamefont {J.~P.}\ \bibnamefont {Olson}},
  \bibinfo {author} {\bibfnamefont {E.~J.}\ \bibnamefont {Rabeaux}}, \bibinfo
  {author} {\bibfnamefont {J.~P.}\ \bibnamefont {Dowling}}, \bibinfo {author}
  {\bibfnamefont {S.~J.}\ \bibnamefont {Olson}},\ and\ \bibinfo {author}
  {\bibfnamefont {P.~P.}\ \bibnamefont {Rohde}},\ }\bibfield  {title} {\bibinfo
  {title} {Linear optical quantum metrology with single photons: Exploiting
  spontaneously generated entanglement to beat the shot-noise limit},\ }\href
  {https://doi.org/10.1103/PhysRevLett.114.170802} {\bibfield  {journal}
  {\bibinfo  {journal} {Phys. Rev. Lett.}\ }\textbf {\bibinfo {volume} {114}},\
  \bibinfo {pages} {170802} (\bibinfo {year} {2015})}\BibitemShut {NoStop}%
\bibitem [{\citenamefont {Olson}\ \emph {et~al.}(2017)\citenamefont {Olson},
  \citenamefont {Motes}, \citenamefont {Birchall}, \citenamefont {Studer},
  \citenamefont {LaBorde}, \citenamefont {Moulder}, \citenamefont {Rohde},\
  and\ \citenamefont {Dowling}}]{Olson2017a}%
  \BibitemOpen
  \bibfield  {author} {\bibinfo {author} {\bibfnamefont {J.~P.}\ \bibnamefont
  {Olson}}, \bibinfo {author} {\bibfnamefont {K.~R.}\ \bibnamefont {Motes}},
  \bibinfo {author} {\bibfnamefont {P.~M.}\ \bibnamefont {Birchall}}, \bibinfo
  {author} {\bibfnamefont {N.~M.}\ \bibnamefont {Studer}}, \bibinfo {author}
  {\bibfnamefont {M.}~\bibnamefont {LaBorde}}, \bibinfo {author} {\bibfnamefont
  {T.}~\bibnamefont {Moulder}}, \bibinfo {author} {\bibfnamefont {P.~P.}\
  \bibnamefont {Rohde}},\ and\ \bibinfo {author} {\bibfnamefont {J.~P.}\
  \bibnamefont {Dowling}},\ }\bibfield  {title} {\bibinfo {title} {Linear
  optical quantum metrology with single photons: Experimental errors, resource
  counting, and quantum cram\'er-rao bounds},\ }\href
  {https://doi.org/10.1103/PhysRevA.96.013810} {\bibfield  {journal} {\bibinfo
  {journal} {Phys. Rev. A}\ }\textbf {\bibinfo {volume} {96}},\ \bibinfo
  {pages} {013810} (\bibinfo {year} {2017})}\BibitemShut {NoStop}%
\bibitem [{\citenamefont {Shen}\ \emph {et~al.}(2017)\citenamefont {Shen},
  \citenamefont {Harris}, \citenamefont {Skirlo}, \citenamefont {Prabhu},
  \citenamefont {Baehr-Jones}, \citenamefont {Hochberg}, \citenamefont {Sun},
  \citenamefont {Zhao}, \citenamefont {Larochelle}, \citenamefont {Englund},\
  and\ \citenamefont {et~al.}}]{Shen2017}%
  \BibitemOpen
  \bibfield  {author} {\bibinfo {author} {\bibfnamefont {Y.}~\bibnamefont
  {Shen}}, \bibinfo {author} {\bibfnamefont {N.~C.}\ \bibnamefont {Harris}},
  \bibinfo {author} {\bibfnamefont {S.}~\bibnamefont {Skirlo}}, \bibinfo
  {author} {\bibfnamefont {M.}~\bibnamefont {Prabhu}}, \bibinfo {author}
  {\bibfnamefont {T.}~\bibnamefont {Baehr-Jones}}, \bibinfo {author}
  {\bibfnamefont {M.}~\bibnamefont {Hochberg}}, \bibinfo {author}
  {\bibfnamefont {X.}~\bibnamefont {Sun}}, \bibinfo {author} {\bibfnamefont
  {S.}~\bibnamefont {Zhao}}, \bibinfo {author} {\bibfnamefont {H.}~\bibnamefont
  {Larochelle}}, \bibinfo {author} {\bibfnamefont {D.}~\bibnamefont
  {Englund}},\ and\ \bibinfo {author} {\bibnamefont {et~al.}},\ }\bibfield
  {title} {\bibinfo {title} {Deep learning with coherent nanophotonic
  circuits},\ }\href {https://doi.org/10.1038/nphoton.2017.93} {\bibfield
  {journal} {\bibinfo  {journal} {Nat. Photonics}\ }\textbf {\bibinfo {volume}
  {11}},\ \bibinfo {pages} {441} (\bibinfo {year} {2017})}\BibitemShut
  {NoStop}%
\bibitem [{\citenamefont {Steinbrecher}\ \emph {et~al.}(2019)\citenamefont
  {Steinbrecher}, \citenamefont {Olson}, \citenamefont {Englund},\ and\
  \citenamefont {Carolan}}]{Steinbrecher2019}%
  \BibitemOpen
  \bibfield  {author} {\bibinfo {author} {\bibfnamefont {G.~R.}\ \bibnamefont
  {Steinbrecher}}, \bibinfo {author} {\bibfnamefont {J.~P.}\ \bibnamefont
  {Olson}}, \bibinfo {author} {\bibfnamefont {D.}~\bibnamefont {Englund}},\
  and\ \bibinfo {author} {\bibfnamefont {J.}~\bibnamefont {Carolan}},\
  }\bibfield  {title} {\bibinfo {title} {Quantum optical neural networks},\
  }\href {https://doi.org/10.1038/s41534-019-0174-7} {\bibfield  {journal}
  {\bibinfo  {journal} {npj Quantum Inf.}\ }\textbf {\bibinfo {volume} {5}},\
  \bibinfo {pages} {60} (\bibinfo {year} {2019})}\BibitemShut {NoStop}%
\bibitem [{\citenamefont {Harris}\ \emph {et~al.}(2020)\citenamefont {Harris},
  \citenamefont {Braid}, \citenamefont {Bunandar}, \citenamefont {Carr},
  \citenamefont {Dobbie}, \citenamefont {Dorta-Quinones}, \citenamefont
  {Elmhurst}, \citenamefont {Forsythe}, \citenamefont {Gould}, \citenamefont
  {Gupta} \emph {et~al.}}]{Harris2020}%
  \BibitemOpen
  \bibfield  {author} {\bibinfo {author} {\bibfnamefont {N.~C.}\ \bibnamefont
  {Harris}}, \bibinfo {author} {\bibfnamefont {R.}~\bibnamefont {Braid}},
  \bibinfo {author} {\bibfnamefont {D.}~\bibnamefont {Bunandar}}, \bibinfo
  {author} {\bibfnamefont {J.}~\bibnamefont {Carr}}, \bibinfo {author}
  {\bibfnamefont {B.}~\bibnamefont {Dobbie}}, \bibinfo {author} {\bibfnamefont
  {C.}~\bibnamefont {Dorta-Quinones}}, \bibinfo {author} {\bibfnamefont
  {J.}~\bibnamefont {Elmhurst}}, \bibinfo {author} {\bibfnamefont
  {M.}~\bibnamefont {Forsythe}}, \bibinfo {author} {\bibfnamefont
  {M.}~\bibnamefont {Gould}}, \bibinfo {author} {\bibfnamefont
  {S.}~\bibnamefont {Gupta}}, \emph {et~al.},\ }\bibfield  {title} {\bibinfo
  {title} {Accelerating artificial intelligence with silicon photonics},\ }in\
  \href {https://doi.org/10.1364/OFC.2020.W3A.3} {\emph {\bibinfo {booktitle}
  {2020 Optical Fiber Communications Conference and Exhibition (OFC)}}}\
  (\bibinfo {organization} {IEEE},\ \bibinfo {year} {2020})\ pp.\ \bibinfo
  {pages} {1--4}\BibitemShut {NoStop}%
\bibitem [{\citenamefont {Capmany}\ and\ \citenamefont
  {P{\'e}rez}(2020)}]{capmany2020programmable}%
  \BibitemOpen
  \bibfield  {author} {\bibinfo {author} {\bibfnamefont {J.}~\bibnamefont
  {Capmany}}\ and\ \bibinfo {author} {\bibfnamefont {D.}~\bibnamefont
  {P{\'e}rez}},\ }\href {https://doi.org/10.1364/IPRSN.2018.ITh1I.2} {\emph
  {\bibinfo {title} {Programmable integrated photonics}}}\ (\bibinfo
  {publisher} {Oxford University Press},\ \bibinfo {year} {2020})\BibitemShut
  {NoStop}%
\bibitem [{\citenamefont {Harris}\ \emph {et~al.}(2018)\citenamefont {Harris},
  \citenamefont {Carolan}, \citenamefont {Bunandar}, \citenamefont {Prabhu},
  \citenamefont {Hochberg}, \citenamefont {Baehr-Jones}, \citenamefont {Fanto},
  \citenamefont {Smith}, \citenamefont {Tison}, \citenamefont {Alsing},\ and\
  \citenamefont {Englund}}]{harris2018linear}%
  \BibitemOpen
  \bibfield  {author} {\bibinfo {author} {\bibfnamefont {N.~C.}\ \bibnamefont
  {Harris}}, \bibinfo {author} {\bibfnamefont {J.}~\bibnamefont {Carolan}},
  \bibinfo {author} {\bibfnamefont {D.}~\bibnamefont {Bunandar}}, \bibinfo
  {author} {\bibfnamefont {M.}~\bibnamefont {Prabhu}}, \bibinfo {author}
  {\bibfnamefont {M.}~\bibnamefont {Hochberg}}, \bibinfo {author}
  {\bibfnamefont {T.}~\bibnamefont {Baehr-Jones}}, \bibinfo {author}
  {\bibfnamefont {M.~L.}\ \bibnamefont {Fanto}}, \bibinfo {author}
  {\bibfnamefont {A.~M.}\ \bibnamefont {Smith}}, \bibinfo {author}
  {\bibfnamefont {C.~C.}\ \bibnamefont {Tison}}, \bibinfo {author}
  {\bibfnamefont {P.~M.}\ \bibnamefont {Alsing}},\ and\ \bibinfo {author}
  {\bibfnamefont {D.}~\bibnamefont {Englund}},\ }\bibfield  {title} {\bibinfo
  {title} {Linear programmable nanophotonic processors},\ }\href
  {https://doi.org/10.1364/OPTICA.5.001623} {\bibfield  {journal} {\bibinfo
  {journal} {Optica}\ }\textbf {\bibinfo {volume} {5}},\ \bibinfo {pages}
  {1623} (\bibinfo {year} {2018})}\BibitemShut {NoStop}%
\bibitem [{\citenamefont {Taballione}\ \emph {et~al.}(2018)\citenamefont
  {Taballione}, \citenamefont {Wolterink}, \citenamefont {Lugani},
  \citenamefont {Eckstein}, \citenamefont {Bell}, \citenamefont {Grootjans},
  \citenamefont {Visscher}, \citenamefont {Renema}, \citenamefont {Geskus},
  \citenamefont {Roeloffzen} \emph {et~al.}}]{taballione20188times}%
  \BibitemOpen
  \bibfield  {author} {\bibinfo {author} {\bibfnamefont {C.}~\bibnamefont
  {Taballione}}, \bibinfo {author} {\bibfnamefont {T.~A.}\ \bibnamefont
  {Wolterink}}, \bibinfo {author} {\bibfnamefont {J.}~\bibnamefont {Lugani}},
  \bibinfo {author} {\bibfnamefont {A.}~\bibnamefont {Eckstein}}, \bibinfo
  {author} {\bibfnamefont {B.~A.}\ \bibnamefont {Bell}}, \bibinfo {author}
  {\bibfnamefont {R.}~\bibnamefont {Grootjans}}, \bibinfo {author}
  {\bibfnamefont {I.}~\bibnamefont {Visscher}}, \bibinfo {author}
  {\bibfnamefont {J.~J.}\ \bibnamefont {Renema}}, \bibinfo {author}
  {\bibfnamefont {D.}~\bibnamefont {Geskus}}, \bibinfo {author} {\bibfnamefont
  {C.~G.}\ \bibnamefont {Roeloffzen}}, \emph {et~al.},\ }\bibfield  {title}
  {\bibinfo {title} {8$\times$ 8 programmable quantum photonic processor based
  on silicon nitride waveguides},\ }in\ \href
  {https://doi.org/10.1364/FIO.2018.JTu3A.58} {\emph {\bibinfo {booktitle}
  {Frontiers in Optics}}}\ (\bibinfo {organization} {Optical Society of
  America},\ \bibinfo {year} {2018})\ pp.\ \bibinfo {pages}
  {JTu3A--58}\BibitemShut {NoStop}%
\bibitem [{\citenamefont {Bogaerts}\ \emph {et~al.}(2020)\citenamefont
  {Bogaerts}, \citenamefont {P{\'e}rez}, \citenamefont {Capmany}, \citenamefont
  {Miller}, \citenamefont {Poon}, \citenamefont {Englund}, \citenamefont
  {Morichetti},\ and\ \citenamefont {Melloni}}]{bogaerts2020programmable}%
  \BibitemOpen
  \bibfield  {author} {\bibinfo {author} {\bibfnamefont {W.}~\bibnamefont
  {Bogaerts}}, \bibinfo {author} {\bibfnamefont {D.}~\bibnamefont {P{\'e}rez}},
  \bibinfo {author} {\bibfnamefont {J.}~\bibnamefont {Capmany}}, \bibinfo
  {author} {\bibfnamefont {D.~A.}\ \bibnamefont {Miller}}, \bibinfo {author}
  {\bibfnamefont {J.}~\bibnamefont {Poon}}, \bibinfo {author} {\bibfnamefont
  {D.}~\bibnamefont {Englund}}, \bibinfo {author} {\bibfnamefont
  {F.}~\bibnamefont {Morichetti}},\ and\ \bibinfo {author} {\bibfnamefont
  {A.}~\bibnamefont {Melloni}},\ }\bibfield  {title} {\bibinfo {title}
  {Programmable photonic circuits},\ }\href
  {https://doi.org/10.1038/s41586-020-2764-0} {\bibfield  {journal} {\bibinfo
  {journal} {Nature}\ }\textbf {\bibinfo {volume} {586}},\ \bibinfo {pages}
  {207} (\bibinfo {year} {2020})}\BibitemShut {NoStop}%
\bibitem [{\citenamefont {Van~Campenhout}\ \emph {et~al.}(2010)\citenamefont
  {Van~Campenhout}, \citenamefont {Green}, \citenamefont {Assefa},\ and\
  \citenamefont {Vlasov}}]{van2010integrated}%
  \BibitemOpen
  \bibfield  {author} {\bibinfo {author} {\bibfnamefont {J.}~\bibnamefont
  {Van~Campenhout}}, \bibinfo {author} {\bibfnamefont {W.~M.}\ \bibnamefont
  {Green}}, \bibinfo {author} {\bibfnamefont {S.}~\bibnamefont {Assefa}},\ and\
  \bibinfo {author} {\bibfnamefont {Y.~A.}\ \bibnamefont {Vlasov}},\ }\bibfield
   {title} {\bibinfo {title} {Integrated nisi waveguide heaters for
  cmos-compatible silicon thermo-optic devices},\ }\href
  {https://doi.org/10.1364/OL.35.001013} {\bibfield  {journal} {\bibinfo
  {journal} {Opt. Lett.}\ }\textbf {\bibinfo {volume} {35}},\ \bibinfo {pages}
  {1013} (\bibinfo {year} {2010})}\BibitemShut {NoStop}%
\bibitem [{\citenamefont {Masood}\ \emph {et~al.}(2013)\citenamefont {Masood},
  \citenamefont {Pantouvaki}, \citenamefont {Lepage}, \citenamefont {Verheyen},
  \citenamefont {Van~Campenhout}, \citenamefont {Absil}, \citenamefont
  {Van~Thourhout},\ and\ \citenamefont {Bogaerts}}]{masood2013comparison}%
  \BibitemOpen
  \bibfield  {author} {\bibinfo {author} {\bibfnamefont {A.}~\bibnamefont
  {Masood}}, \bibinfo {author} {\bibfnamefont {M.}~\bibnamefont {Pantouvaki}},
  \bibinfo {author} {\bibfnamefont {G.}~\bibnamefont {Lepage}}, \bibinfo
  {author} {\bibfnamefont {P.}~\bibnamefont {Verheyen}}, \bibinfo {author}
  {\bibfnamefont {J.}~\bibnamefont {Van~Campenhout}}, \bibinfo {author}
  {\bibfnamefont {P.}~\bibnamefont {Absil}}, \bibinfo {author} {\bibfnamefont
  {D.}~\bibnamefont {Van~Thourhout}},\ and\ \bibinfo {author} {\bibfnamefont
  {W.}~\bibnamefont {Bogaerts}},\ }\bibfield  {title} {\bibinfo {title}
  {Comparison of heater architectures for thermal control of silicon photonic
  circuits},\ }in\ \href@noop {} {\emph {\bibinfo {booktitle} {10th
  International Conference on Group IV Photonics}}}\ (\bibinfo {organization}
  {IEEE},\ \bibinfo {year} {2013})\ pp.\ \bibinfo {pages} {83--84}\BibitemShut
  {NoStop}%
\bibitem [{\citenamefont {Soref}\ and\ \citenamefont
  {Bennett}(1987)}]{soref1987electrooptical}%
  \BibitemOpen
  \bibfield  {author} {\bibinfo {author} {\bibfnamefont {R.}~\bibnamefont
  {Soref}}\ and\ \bibinfo {author} {\bibfnamefont {B.}~\bibnamefont
  {Bennett}},\ }\bibfield  {title} {\bibinfo {title} {Electrooptical effects in
  silicon},\ }\href {https://doi.org/10.1109/JQE.1987.1073206} {\bibfield
  {journal} {\bibinfo  {journal} {IEEE J. Quantum Electron.}\ }\textbf
  {\bibinfo {volume} {23}},\ \bibinfo {pages} {123} (\bibinfo {year}
  {1987})}\BibitemShut {NoStop}%
\bibitem [{\citenamefont {Reed}\ \emph {et~al.}(2010)\citenamefont {Reed},
  \citenamefont {Mashanovich}, \citenamefont {Gardes},\ and\ \citenamefont
  {Thomson}}]{reed2010silicon}%
  \BibitemOpen
  \bibfield  {author} {\bibinfo {author} {\bibfnamefont {G.~T.}\ \bibnamefont
  {Reed}}, \bibinfo {author} {\bibfnamefont {G.}~\bibnamefont {Mashanovich}},
  \bibinfo {author} {\bibfnamefont {F.~Y.}\ \bibnamefont {Gardes}},\ and\
  \bibinfo {author} {\bibfnamefont {D.}~\bibnamefont {Thomson}},\ }\bibfield
  {title} {\bibinfo {title} {Silicon optical modulators},\ }\href
  {https://doi.org/10.1038/nphoton.2010.179} {\bibfield  {journal} {\bibinfo
  {journal} {Nat. Photonics}\ }\textbf {\bibinfo {volume} {4}},\ \bibinfo
  {pages} {518} (\bibinfo {year} {2010})}\BibitemShut {NoStop}%
\bibitem [{\citenamefont {Harris}\ \emph {et~al.}(2017)\citenamefont {Harris},
  \citenamefont {Steinbrecher}, \citenamefont {Prabhu}, \citenamefont {Lahini},
  \citenamefont {Mower}, \citenamefont {Bunandar}, \citenamefont {Chen},
  \citenamefont {Wong}, \citenamefont {Baehr-Jones}, \citenamefont {Hochberg},
  \citenamefont {Lloyd},\ and\ \citenamefont {Englund}}]{harris2017quantum}%
  \BibitemOpen
  \bibfield  {author} {\bibinfo {author} {\bibfnamefont {N.~C.}\ \bibnamefont
  {Harris}}, \bibinfo {author} {\bibfnamefont {G.~R.}\ \bibnamefont
  {Steinbrecher}}, \bibinfo {author} {\bibfnamefont {M.}~\bibnamefont
  {Prabhu}}, \bibinfo {author} {\bibfnamefont {Y.}~\bibnamefont {Lahini}},
  \bibinfo {author} {\bibfnamefont {J.}~\bibnamefont {Mower}}, \bibinfo
  {author} {\bibfnamefont {D.}~\bibnamefont {Bunandar}}, \bibinfo {author}
  {\bibfnamefont {C.}~\bibnamefont {Chen}}, \bibinfo {author} {\bibfnamefont
  {F.~N.~C.}\ \bibnamefont {Wong}}, \bibinfo {author} {\bibfnamefont
  {T.}~\bibnamefont {Baehr-Jones}}, \bibinfo {author} {\bibfnamefont
  {M.}~\bibnamefont {Hochberg}}, \bibinfo {author} {\bibfnamefont
  {S.}~\bibnamefont {Lloyd}},\ and\ \bibinfo {author} {\bibfnamefont
  {D.}~\bibnamefont {Englund}},\ }\bibfield  {title} {\bibinfo {title} {Quantum
  transport simulations in a programmable nanophotonic processor},\ }\href
  {https://doi.org/10.1038/nphoton.2017.95} {\bibfield  {journal} {\bibinfo
  {journal} {Nat. Photonics}\ }\textbf {\bibinfo {volume} {11}},\ \bibinfo
  {pages} {447} (\bibinfo {year} {2017})}\BibitemShut {NoStop}%
\bibitem [{\citenamefont {Suzuki}\ \emph {et~al.}(2018)\citenamefont {Suzuki},
  \citenamefont {Konoike}, \citenamefont {Hasegawa}, \citenamefont {Suda},
  \citenamefont {Matsuura}, \citenamefont {Ikeda}, \citenamefont {Namiki},\
  and\ \citenamefont {Kawashima}}]{suzuki2018low}%
  \BibitemOpen
  \bibfield  {author} {\bibinfo {author} {\bibfnamefont {K.}~\bibnamefont
  {Suzuki}}, \bibinfo {author} {\bibfnamefont {R.}~\bibnamefont {Konoike}},
  \bibinfo {author} {\bibfnamefont {J.}~\bibnamefont {Hasegawa}}, \bibinfo
  {author} {\bibfnamefont {S.}~\bibnamefont {Suda}}, \bibinfo {author}
  {\bibfnamefont {H.}~\bibnamefont {Matsuura}}, \bibinfo {author}
  {\bibfnamefont {K.}~\bibnamefont {Ikeda}}, \bibinfo {author} {\bibfnamefont
  {S.}~\bibnamefont {Namiki}},\ and\ \bibinfo {author} {\bibfnamefont
  {H.}~\bibnamefont {Kawashima}},\ }\bibfield  {title} {\bibinfo {title}
  {{Low-insertion-loss and power-efficient 32$\times$ 32 silicon photonics
  switch with extremely high-$\Delta$ silica PLC connector}},\ }\href
  {https://doi.org/10.1109/JLT.2018.2867575} {\bibfield  {journal} {\bibinfo
  {journal} {J. Light. Technol.}\ }\textbf {\bibinfo {volume} {37}},\ \bibinfo
  {pages} {116} (\bibinfo {year} {2018})}\BibitemShut {NoStop}%
\bibitem [{\citenamefont {Taballione}\ \emph {et~al.}(2020)\citenamefont
  {Taballione}, \citenamefont {van~der Meer}, \citenamefont {Snijders},
  \citenamefont {Hooijschuur}, \citenamefont {Epping}, \citenamefont
  {de~Goede}, \citenamefont {Kassenberg}, \citenamefont {Venderbosch},
  \citenamefont {Toebes}, \citenamefont {Vlekkert} \emph
  {et~al.}}]{taballione202012}%
  \BibitemOpen
  \bibfield  {author} {\bibinfo {author} {\bibfnamefont {C.}~\bibnamefont
  {Taballione}}, \bibinfo {author} {\bibfnamefont {R.}~\bibnamefont {van~der
  Meer}}, \bibinfo {author} {\bibfnamefont {H.~J.}\ \bibnamefont {Snijders}},
  \bibinfo {author} {\bibfnamefont {P.}~\bibnamefont {Hooijschuur}}, \bibinfo
  {author} {\bibfnamefont {J.~P.}\ \bibnamefont {Epping}}, \bibinfo {author}
  {\bibfnamefont {M.}~\bibnamefont {de~Goede}}, \bibinfo {author}
  {\bibfnamefont {B.}~\bibnamefont {Kassenberg}}, \bibinfo {author}
  {\bibfnamefont {P.}~\bibnamefont {Venderbosch}}, \bibinfo {author}
  {\bibfnamefont {C.}~\bibnamefont {Toebes}}, \bibinfo {author} {\bibfnamefont
  {H.~v.~d.}\ \bibnamefont {Vlekkert}}, \emph {et~al.},\ }\bibfield  {title}
  {\bibinfo {title} {A 12-mode universal photonic processor for quantum
  information processing},\ }\href@noop {} {\  (\bibinfo {year} {2020})},\
  \Eprint {https://arxiv.org/abs/2012.05673} {arXiv:2012.05673 [quant-ph]}
  \BibitemShut {NoStop}%
\bibitem [{\citenamefont {Clements}\ \emph {et~al.}(2016)\citenamefont
  {Clements}, \citenamefont {Humphreys}, \citenamefont {Metcalf}, \citenamefont
  {Kolthammer},\ and\ \citenamefont {Walsmley}}]{clements2016optimal}%
  \BibitemOpen
  \bibfield  {author} {\bibinfo {author} {\bibfnamefont {W.~R.}\ \bibnamefont
  {Clements}}, \bibinfo {author} {\bibfnamefont {P.~C.}\ \bibnamefont
  {Humphreys}}, \bibinfo {author} {\bibfnamefont {B.~J.}\ \bibnamefont
  {Metcalf}}, \bibinfo {author} {\bibfnamefont {W.~S.}\ \bibnamefont
  {Kolthammer}},\ and\ \bibinfo {author} {\bibfnamefont {I.~A.}\ \bibnamefont
  {Walsmley}},\ }\bibfield  {title} {\bibinfo {title} {Optimal design for
  universal multiport interferometers},\ }\href
  {https://doi.org/10.1364/OPTICA.3.001460} {\bibfield  {journal} {\bibinfo
  {journal} {Optica}\ }\textbf {\bibinfo {volume} {3}},\ \bibinfo {pages}
  {1460} (\bibinfo {year} {2016})},\ \Eprint {https://arxiv.org/abs/1603.08788}
  {arXiv:1603.08788} \BibitemShut {NoStop}%
\bibitem [{\citenamefont {Reck}\ \emph {et~al.}(1994)\citenamefont {Reck},
  \citenamefont {Zeilinger}, \citenamefont {Bernstein},\ and\ \citenamefont
  {Bertani}}]{reck1994experimental}%
  \BibitemOpen
  \bibfield  {author} {\bibinfo {author} {\bibfnamefont {M.}~\bibnamefont
  {Reck}}, \bibinfo {author} {\bibfnamefont {A.}~\bibnamefont {Zeilinger}},
  \bibinfo {author} {\bibfnamefont {H.~J.}\ \bibnamefont {Bernstein}},\ and\
  \bibinfo {author} {\bibfnamefont {P.}~\bibnamefont {Bertani}},\ }\bibfield
  {title} {\bibinfo {title} {{E}xperimental realization of any discrete unitary
  operator},\ }\href {https://doi.org/10.1103/PhysRevLett.73.58} {\bibfield
  {journal} {\bibinfo  {journal} {Phys. Rev. Lett.}\ }\textbf {\bibinfo
  {volume} {73}},\ \bibinfo {pages} {58} (\bibinfo {year} {1994})}\BibitemShut
  {NoStop}%
\bibitem [{\citenamefont {de~Guise}\ \emph {et~al.}(2018)\citenamefont
  {de~Guise}, \citenamefont {Di~Matteo},\ and\ \citenamefont
  {S\'anchez-Soto}}]{guise2018simple}%
  \BibitemOpen
  \bibfield  {author} {\bibinfo {author} {\bibfnamefont {H.}~\bibnamefont
  {de~Guise}}, \bibinfo {author} {\bibfnamefont {O.}~\bibnamefont
  {Di~Matteo}},\ and\ \bibinfo {author} {\bibfnamefont {L.~L.}\ \bibnamefont
  {S\'anchez-Soto}},\ }\bibfield  {title} {\bibinfo {title} {Simple
  factorization of unitary transformations},\ }\href
  {https://doi.org/10.1103/PhysRevA.97.022328} {\bibfield  {journal} {\bibinfo
  {journal} {Phys. Rev. A}\ }\textbf {\bibinfo {volume} {97}},\ \bibinfo
  {pages} {022328} (\bibinfo {year} {2018})},\ \Eprint
  {https://arxiv.org/abs/1708.00735} {arXiv:1708.00735} \BibitemShut {NoStop}%
\bibitem [{\citenamefont {Thiele}\ \emph {et~al.}(2020)\citenamefont {Thiele},
  \citenamefont {Vom~Bruch}, \citenamefont {Quiring}, \citenamefont {Ricken},
  \citenamefont {Herrmann}, \citenamefont {Eigner}, \citenamefont
  {Silberhorn},\ and\ \citenamefont {Bartley}}]{thiele2020cryogenic}%
  \BibitemOpen
  \bibfield  {author} {\bibinfo {author} {\bibfnamefont {F.}~\bibnamefont
  {Thiele}}, \bibinfo {author} {\bibfnamefont {F.}~\bibnamefont {Vom~Bruch}},
  \bibinfo {author} {\bibfnamefont {V.}~\bibnamefont {Quiring}}, \bibinfo
  {author} {\bibfnamefont {R.}~\bibnamefont {Ricken}}, \bibinfo {author}
  {\bibfnamefont {H.}~\bibnamefont {Herrmann}}, \bibinfo {author}
  {\bibfnamefont {C.}~\bibnamefont {Eigner}}, \bibinfo {author} {\bibfnamefont
  {C.}~\bibnamefont {Silberhorn}},\ and\ \bibinfo {author} {\bibfnamefont
  {T.~J.}\ \bibnamefont {Bartley}},\ }\bibfield  {title} {\bibinfo {title}
  {Cryogenic electro-optic polarisation conversion in titanium in-diffused
  lithium niobate waveguides},\ }\href {https://doi.org/10.1364/OE.399818}
  {\bibfield  {journal} {\bibinfo  {journal} {Opt. Express}\ }\textbf {\bibinfo
  {volume} {28}},\ \bibinfo {pages} {28961} (\bibinfo {year}
  {2020})}\BibitemShut {NoStop}%
\bibitem [{\citenamefont {Eltes}\ \emph {et~al.}(2020)\citenamefont {Eltes},
  \citenamefont {Villarreal-Garcia}, \citenamefont {Caimi}, \citenamefont
  {Siegwart}, \citenamefont {Gentile}, \citenamefont {Hart}, \citenamefont
  {Stark}, \citenamefont {Marshall}, \citenamefont {Thompson}, \citenamefont
  {Barreto} \emph {et~al.}}]{eltes2020an-integrated}%
  \BibitemOpen
  \bibfield  {author} {\bibinfo {author} {\bibfnamefont {F.}~\bibnamefont
  {Eltes}}, \bibinfo {author} {\bibfnamefont {G.~E.}\ \bibnamefont
  {Villarreal-Garcia}}, \bibinfo {author} {\bibfnamefont {D.}~\bibnamefont
  {Caimi}}, \bibinfo {author} {\bibfnamefont {H.}~\bibnamefont {Siegwart}},
  \bibinfo {author} {\bibfnamefont {A.~A.}\ \bibnamefont {Gentile}}, \bibinfo
  {author} {\bibfnamefont {A.}~\bibnamefont {Hart}}, \bibinfo {author}
  {\bibfnamefont {P.}~\bibnamefont {Stark}}, \bibinfo {author} {\bibfnamefont
  {G.~D.}\ \bibnamefont {Marshall}}, \bibinfo {author} {\bibfnamefont {M.~G.}\
  \bibnamefont {Thompson}}, \bibinfo {author} {\bibfnamefont {J.}~\bibnamefont
  {Barreto}}, \emph {et~al.},\ }\bibfield  {title} {\bibinfo {title} {An
  integrated optical modulator operating at cryogenic temperatures},\ }\href
  {https://doi.org/10.1038/s41563-020-0725-5} {\bibfield  {journal} {\bibinfo
  {journal} {Nat. Mater.}\ }\textbf {\bibinfo {volume} {19}},\ \bibinfo {pages}
  {1164} (\bibinfo {year} {2020})}\BibitemShut {NoStop}%
\bibitem [{\citenamefont {Vaidya}\ \emph {et~al.}(2020)\citenamefont {Vaidya},
  \citenamefont {Morrison}, \citenamefont {Helt}, \citenamefont {Shahrokshahi},
  \citenamefont {Mahler}, \citenamefont {Collins}, \citenamefont {Tan},
  \citenamefont {Lavoie}, \citenamefont {Repingon}, \citenamefont {Menotti}
  \emph {et~al.}}]{vaidya2020broadband}%
  \BibitemOpen
  \bibfield  {author} {\bibinfo {author} {\bibfnamefont {V.~D.}\ \bibnamefont
  {Vaidya}}, \bibinfo {author} {\bibfnamefont {B.}~\bibnamefont {Morrison}},
  \bibinfo {author} {\bibfnamefont {L.}~\bibnamefont {Helt}}, \bibinfo {author}
  {\bibfnamefont {R.}~\bibnamefont {Shahrokshahi}}, \bibinfo {author}
  {\bibfnamefont {D.}~\bibnamefont {Mahler}}, \bibinfo {author} {\bibfnamefont
  {M.}~\bibnamefont {Collins}}, \bibinfo {author} {\bibfnamefont
  {K.}~\bibnamefont {Tan}}, \bibinfo {author} {\bibfnamefont {J.}~\bibnamefont
  {Lavoie}}, \bibinfo {author} {\bibfnamefont {A.}~\bibnamefont {Repingon}},
  \bibinfo {author} {\bibfnamefont {M.}~\bibnamefont {Menotti}}, \emph
  {et~al.},\ }\bibfield  {title} {\bibinfo {title} {Broadband
  quadrature-squeezed vacuum and nonclassical photon number correlations from a
  nanophotonic device},\ }\href {https://doi.org/10.1126/sciadv.aba9186}
  {\bibfield  {journal} {\bibinfo  {journal} {Sci. Adv.}\ }\textbf {\bibinfo
  {volume} {6}},\ \bibinfo {pages} {eaba9186} (\bibinfo {year}
  {2020})}\BibitemShut {NoStop}%
\bibitem [{\citenamefont {Milanizadeh}\ \emph {et~al.}(2020)\citenamefont
  {Milanizadeh}, \citenamefont {Ahmadi}, \citenamefont {Petrini}, \citenamefont
  {Aguiar}, \citenamefont {Mazzanti}, \citenamefont {Zanetto}, \citenamefont
  {Guglielmi}, \citenamefont {Sampietro}, \citenamefont {Morichetti},\ and\
  \citenamefont {Melloni}}]{milanizadeh2020control}%
  \BibitemOpen
  \bibfield  {author} {\bibinfo {author} {\bibfnamefont {M.}~\bibnamefont
  {Milanizadeh}}, \bibinfo {author} {\bibfnamefont {S.}~\bibnamefont {Ahmadi}},
  \bibinfo {author} {\bibfnamefont {M.}~\bibnamefont {Petrini}}, \bibinfo
  {author} {\bibfnamefont {D.}~\bibnamefont {Aguiar}}, \bibinfo {author}
  {\bibfnamefont {R.}~\bibnamefont {Mazzanti}}, \bibinfo {author}
  {\bibfnamefont {F.}~\bibnamefont {Zanetto}}, \bibinfo {author} {\bibfnamefont
  {E.}~\bibnamefont {Guglielmi}}, \bibinfo {author} {\bibfnamefont
  {M.}~\bibnamefont {Sampietro}}, \bibinfo {author} {\bibfnamefont
  {F.}~\bibnamefont {Morichetti}},\ and\ \bibinfo {author} {\bibfnamefont
  {A.}~\bibnamefont {Melloni}},\ }\bibfield  {title} {\bibinfo {title} {Control
  and calibration recipes for photonic integrated circuits},\ }\href
  {https://doi.org/10.1109/JSTQE.2020.2975657} {\bibfield  {journal} {\bibinfo
  {journal} {IEEE Journal of Selected Topics in Quantum Electronics}\ }\textbf
  {\bibinfo {volume} {26}},\ \bibinfo {pages} {1} (\bibinfo {year}
  {2020})}\BibitemShut {NoStop}%
\bibitem [{\citenamefont {Fatemi}\ \emph {et~al.}(2019)\citenamefont {Fatemi},
  \citenamefont {Khachaturian},\ and\ \citenamefont
  {Hajimiri}}]{fatemi2019nonuniform}%
  \BibitemOpen
  \bibfield  {author} {\bibinfo {author} {\bibfnamefont {R.}~\bibnamefont
  {Fatemi}}, \bibinfo {author} {\bibfnamefont {A.}~\bibnamefont
  {Khachaturian}},\ and\ \bibinfo {author} {\bibfnamefont {A.}~\bibnamefont
  {Hajimiri}},\ }\bibfield  {title} {\bibinfo {title} {A nonuniform sparse 2-d
  large-fov optical phased array with a low-power pwm drive},\ }\href
  {https://doi.org/10.1109/JSSC.2019.2896767} {\bibfield  {journal} {\bibinfo
  {journal} {IEEE Journal of Solid-State Circuits}\ }\textbf {\bibinfo {volume}
  {54}},\ \bibinfo {pages} {1200} (\bibinfo {year} {2019})}\BibitemShut
  {NoStop}%
\bibitem [{\citenamefont {Su}\ \emph {et~al.}(2020)\citenamefont {Su},
  \citenamefont {Israel}, \citenamefont {Sharma}, \citenamefont {Qi},
  \citenamefont {Dhand},\ and\ \citenamefont {Br{\'a}dler}}]{su2020error}%
  \BibitemOpen
  \bibfield  {author} {\bibinfo {author} {\bibfnamefont {D.}~\bibnamefont
  {Su}}, \bibinfo {author} {\bibfnamefont {R.}~\bibnamefont {Israel}}, \bibinfo
  {author} {\bibfnamefont {K.}~\bibnamefont {Sharma}}, \bibinfo {author}
  {\bibfnamefont {H.}~\bibnamefont {Qi}}, \bibinfo {author} {\bibfnamefont
  {I.}~\bibnamefont {Dhand}},\ and\ \bibinfo {author} {\bibfnamefont
  {K.}~\bibnamefont {Br{\'a}dler}},\ }\bibfield  {title} {\bibinfo {title}
  {Error mitigation on a near-term quantum photonic device},\ }\href@noop {} {\
   (\bibinfo {year} {2020})},\ \Eprint {https://arxiv.org/abs/2008.06670}
  {arXiv:2008.06670 [quant-ph]} \BibitemShut {NoStop}%
\bibitem [{\citenamefont {Dhand}(2019)}]{dhand2019circumventing}%
  \BibitemOpen
  \bibfield  {author} {\bibinfo {author} {\bibfnamefont {I.}~\bibnamefont
  {Dhand}},\ }\bibfield  {title} {\bibinfo {title} {Circumventing defective
  components in linear optical interferometers},\ }\href@noop {} {\  (\bibinfo
  {year} {2019})},\ \Eprint {https://arxiv.org/abs/1912.08789}
  {arXiv:1912.08789 [quant-ph]} \BibitemShut {NoStop}%
\bibitem [{\citenamefont {Crespi}\ \emph {et~al.}(2013)\citenamefont {Crespi},
  \citenamefont {Osellame}, \citenamefont {Ramponi}, \citenamefont {Brod},
  \citenamefont {Galvao}, \citenamefont {Spagnolo}, \citenamefont {Vitelli},
  \citenamefont {Maiorino}, \citenamefont {Mataloni},\ and\ \citenamefont
  {Sciarrino}}]{crespi2013integrated}%
  \BibitemOpen
  \bibfield  {author} {\bibinfo {author} {\bibfnamefont {A.}~\bibnamefont
  {Crespi}}, \bibinfo {author} {\bibfnamefont {R.}~\bibnamefont {Osellame}},
  \bibinfo {author} {\bibfnamefont {R.}~\bibnamefont {Ramponi}}, \bibinfo
  {author} {\bibfnamefont {D.~J.}\ \bibnamefont {Brod}}, \bibinfo {author}
  {\bibfnamefont {E.~F.}\ \bibnamefont {Galvao}}, \bibinfo {author}
  {\bibfnamefont {N.}~\bibnamefont {Spagnolo}}, \bibinfo {author}
  {\bibfnamefont {C.}~\bibnamefont {Vitelli}}, \bibinfo {author} {\bibfnamefont
  {E.}~\bibnamefont {Maiorino}}, \bibinfo {author} {\bibfnamefont
  {P.}~\bibnamefont {Mataloni}},\ and\ \bibinfo {author} {\bibfnamefont
  {F.}~\bibnamefont {Sciarrino}},\ }\bibfield  {title} {\bibinfo {title}
  {Integrated multimode interferometers with arbitrary designs for photonic
  boson sampling},\ }\href {https://doi.org/10.1038/nphoton.2013.112}
  {\bibfield  {journal} {\bibinfo  {journal} {Nat. Photonics}\ }\textbf
  {\bibinfo {volume} {7}},\ \bibinfo {pages} {545} (\bibinfo {year}
  {2013})}\BibitemShut {NoStop}%
\bibitem [{\citenamefont {Wang}\ \emph {et~al.}(2019)\citenamefont {Wang},
  \citenamefont {Yang}, \citenamefont {Zhang}, \citenamefont {Miao},
  \citenamefont {Zhao},\ and\ \citenamefont {Xu}}]{wang2019high}%
  \BibitemOpen
  \bibfield  {author} {\bibinfo {author} {\bibfnamefont {L.}~\bibnamefont
  {Wang}}, \bibinfo {author} {\bibfnamefont {L.}~\bibnamefont {Yang}}, \bibinfo
  {author} {\bibfnamefont {C.}~\bibnamefont {Zhang}}, \bibinfo {author}
  {\bibfnamefont {C.}~\bibnamefont {Miao}}, \bibinfo {author} {\bibfnamefont
  {J.}~\bibnamefont {Zhao}},\ and\ \bibinfo {author} {\bibfnamefont
  {W.}~\bibnamefont {Xu}},\ }\bibfield  {title} {\bibinfo {title} {High
  sensitivity and low loss open-cavity mach-zehnder interferometer based on
  multimode interference coupling for refractive index measurement},\ }\href
  {https://doi.org/10.1016/j.optlastec.2018.08.002} {\bibfield  {journal}
  {\bibinfo  {journal} {Opt. Laser. Technol.}\ }\textbf {\bibinfo {volume}
  {109}},\ \bibinfo {pages} {193} (\bibinfo {year} {2019})}\BibitemShut
  {NoStop}%
\bibitem [{\citenamefont {Suzuki}\ \emph {et~al.}(2015)\citenamefont {Suzuki},
  \citenamefont {Cong}, \citenamefont {Tanizawa}, \citenamefont {Kim},
  \citenamefont {Ikeda}, \citenamefont {Namiki},\ and\ \citenamefont
  {Kawashima}}]{suzuki2015ultra-high-extinction-ratio}%
  \BibitemOpen
  \bibfield  {author} {\bibinfo {author} {\bibfnamefont {K.}~\bibnamefont
  {Suzuki}}, \bibinfo {author} {\bibfnamefont {G.}~\bibnamefont {Cong}},
  \bibinfo {author} {\bibfnamefont {K.}~\bibnamefont {Tanizawa}}, \bibinfo
  {author} {\bibfnamefont {S.-H.}\ \bibnamefont {Kim}}, \bibinfo {author}
  {\bibfnamefont {K.}~\bibnamefont {Ikeda}}, \bibinfo {author} {\bibfnamefont
  {S.}~\bibnamefont {Namiki}},\ and\ \bibinfo {author} {\bibfnamefont
  {H.}~\bibnamefont {Kawashima}},\ }\bibfield  {title} {\bibinfo {title}
  {Ultra-high-extinction-ratio 2 {\texttimes} 2 silicon optical switch with
  variable splitter},\ }\href {https://doi.org/10.1364/OE.23.009086} {\bibfield
   {journal} {\bibinfo  {journal} {Opt. Express}\ }\textbf {\bibinfo {volume}
  {23}},\ \bibinfo {pages} {9086} (\bibinfo {year} {2015})}\BibitemShut
  {NoStop}%
\bibitem [{\citenamefont {Mower}\ \emph {et~al.}(2015)\citenamefont {Mower},
  \citenamefont {Harris}, \citenamefont {Steinbrecher}, \citenamefont
  {Lahini},\ and\ \citenamefont {Englund}}]{mower2015high-fidelity}%
  \BibitemOpen
  \bibfield  {author} {\bibinfo {author} {\bibfnamefont {J.}~\bibnamefont
  {Mower}}, \bibinfo {author} {\bibfnamefont {N.~C.}\ \bibnamefont {Harris}},
  \bibinfo {author} {\bibfnamefont {G.~R.}\ \bibnamefont {Steinbrecher}},
  \bibinfo {author} {\bibfnamefont {Y.}~\bibnamefont {Lahini}},\ and\ \bibinfo
  {author} {\bibfnamefont {D.}~\bibnamefont {Englund}},\ }\bibfield  {title}
  {\bibinfo {title} {High-fidelity quantum state evolution in imperfect
  photonic integrated circuits},\ }\href
  {https://doi.org/10.1103/PhysRevA.92.032322} {\bibfield  {journal} {\bibinfo
  {journal} {Phys. Rev. A}\ }\textbf {\bibinfo {volume} {92}},\ \bibinfo
  {pages} {032322} (\bibinfo {year} {2015})}\BibitemShut {NoStop}%
\bibitem [{\citenamefont {Serafini}(2017)}]{book_serafini}%
  \BibitemOpen
  \bibfield  {author} {\bibinfo {author} {\bibfnamefont {A.}~\bibnamefont
  {Serafini}},\ }\href {https://doi.org/10.1201/9781315118727} {\emph {\bibinfo
  {title} {{Quantum Continuous Variables}}}}\ (\bibinfo  {publisher} {CRC
  Press},\ \bibinfo {year} {2017})\BibitemShut {NoStop}%
\bibitem [{\citenamefont {Kok}\ \emph {et~al.}(2007)\citenamefont {Kok},
  \citenamefont {Munro}, \citenamefont {Nemoto}, \citenamefont {Ralph},
  \citenamefont {Dowling},\ and\ \citenamefont {Milburn}}]{kok2007linear}%
  \BibitemOpen
  \bibfield  {author} {\bibinfo {author} {\bibfnamefont {P.}~\bibnamefont
  {Kok}}, \bibinfo {author} {\bibfnamefont {W.~J.}\ \bibnamefont {Munro}},
  \bibinfo {author} {\bibfnamefont {K.}~\bibnamefont {Nemoto}}, \bibinfo
  {author} {\bibfnamefont {T.~C.}\ \bibnamefont {Ralph}}, \bibinfo {author}
  {\bibfnamefont {J.~P.}\ \bibnamefont {Dowling}},\ and\ \bibinfo {author}
  {\bibfnamefont {G.~J.}\ \bibnamefont {Milburn}},\ }\bibfield  {title}
  {\bibinfo {title} {Linear optical quantum computing with photonic qubits},\
  }\href {https://doi.org/10.1103/revmodphys.79.135} {\bibfield  {journal}
  {\bibinfo  {journal} {Rev. Mod. Phys.}\ }\textbf {\bibinfo {volume} {79}},\
  \bibinfo {pages} {135} (\bibinfo {year} {2007})}\BibitemShut {NoStop}%
\bibitem [{\citenamefont {Arrazola}\ \emph {et~al.}(2021)\citenamefont
  {Arrazola}, \citenamefont {Bergholm}, \citenamefont {Br{\'a}dler},
  \citenamefont {Bromley}, \citenamefont {Collins}, \citenamefont {Dhand},
  \citenamefont {Fumagalli}, \citenamefont {Gerrits}, \citenamefont {Goussev},
  \citenamefont {Helt}, \citenamefont {Hundal}, \citenamefont {Isacsson},
  \citenamefont {Israel}, \citenamefont {Izaac}, \citenamefont {Jahangiri},
  \citenamefont {Janik}, \citenamefont {Killoran}, \citenamefont {Kumar},
  \citenamefont {Lavoie}, \citenamefont {Lita}, \citenamefont {Mahler},
  \citenamefont {Menotti}, \citenamefont {Morrison}, \citenamefont {Nam},
  \citenamefont {Neuhaus}, \citenamefont {Qi}, \citenamefont {Quesada},
  \citenamefont {Repingon}, \citenamefont {Sabapathy}, \citenamefont {Schuld},
  \citenamefont {Su}, \citenamefont {Swinarton}, \citenamefont {Sz{\'a}va},
  \citenamefont {Tan}, \citenamefont {Tan}, \citenamefont {Vaidya},
  \citenamefont {Vernon}, \citenamefont {Zabaneh},\ and\ \citenamefont
  {Zhang}}]{arrazola2021quantum}%
  \BibitemOpen
  \bibfield  {author} {\bibinfo {author} {\bibfnamefont {J.~M.}\ \bibnamefont
  {Arrazola}}, \bibinfo {author} {\bibfnamefont {V.}~\bibnamefont {Bergholm}},
  \bibinfo {author} {\bibfnamefont {K.}~\bibnamefont {Br{\'a}dler}}, \bibinfo
  {author} {\bibfnamefont {T.~R.}\ \bibnamefont {Bromley}}, \bibinfo {author}
  {\bibfnamefont {M.~J.}\ \bibnamefont {Collins}}, \bibinfo {author}
  {\bibfnamefont {I.}~\bibnamefont {Dhand}}, \bibinfo {author} {\bibfnamefont
  {A.}~\bibnamefont {Fumagalli}}, \bibinfo {author} {\bibfnamefont
  {T.}~\bibnamefont {Gerrits}}, \bibinfo {author} {\bibfnamefont
  {A.}~\bibnamefont {Goussev}}, \bibinfo {author} {\bibfnamefont {L.~G.}\
  \bibnamefont {Helt}}, \bibinfo {author} {\bibfnamefont {J.}~\bibnamefont
  {Hundal}}, \bibinfo {author} {\bibfnamefont {T.}~\bibnamefont {Isacsson}},
  \bibinfo {author} {\bibfnamefont {R.~B.}\ \bibnamefont {Israel}}, \bibinfo
  {author} {\bibfnamefont {J.}~\bibnamefont {Izaac}}, \bibinfo {author}
  {\bibfnamefont {S.}~\bibnamefont {Jahangiri}}, \bibinfo {author}
  {\bibfnamefont {R.}~\bibnamefont {Janik}}, \bibinfo {author} {\bibfnamefont
  {N.}~\bibnamefont {Killoran}}, \bibinfo {author} {\bibfnamefont {S.~P.}\
  \bibnamefont {Kumar}}, \bibinfo {author} {\bibfnamefont {J.}~\bibnamefont
  {Lavoie}}, \bibinfo {author} {\bibfnamefont {A.~E.}\ \bibnamefont {Lita}},
  \bibinfo {author} {\bibfnamefont {D.~H.}\ \bibnamefont {Mahler}}, \bibinfo
  {author} {\bibfnamefont {M.}~\bibnamefont {Menotti}}, \bibinfo {author}
  {\bibfnamefont {B.}~\bibnamefont {Morrison}}, \bibinfo {author}
  {\bibfnamefont {S.~W.}\ \bibnamefont {Nam}}, \bibinfo {author} {\bibfnamefont
  {L.}~\bibnamefont {Neuhaus}}, \bibinfo {author} {\bibfnamefont {H.~Y.}\
  \bibnamefont {Qi}}, \bibinfo {author} {\bibfnamefont {N.}~\bibnamefont
  {Quesada}}, \bibinfo {author} {\bibfnamefont {A.}~\bibnamefont {Repingon}},
  \bibinfo {author} {\bibfnamefont {K.~K.}\ \bibnamefont {Sabapathy}}, \bibinfo
  {author} {\bibfnamefont {M.}~\bibnamefont {Schuld}}, \bibinfo {author}
  {\bibfnamefont {D.}~\bibnamefont {Su}}, \bibinfo {author} {\bibfnamefont
  {J.}~\bibnamefont {Swinarton}}, \bibinfo {author} {\bibfnamefont
  {A.}~\bibnamefont {Sz{\'a}va}}, \bibinfo {author} {\bibfnamefont
  {K.}~\bibnamefont {Tan}}, \bibinfo {author} {\bibfnamefont {P.}~\bibnamefont
  {Tan}}, \bibinfo {author} {\bibfnamefont {V.~D.}\ \bibnamefont {Vaidya}},
  \bibinfo {author} {\bibfnamefont {Z.}~\bibnamefont {Vernon}}, \bibinfo
  {author} {\bibfnamefont {Z.}~\bibnamefont {Zabaneh}},\ and\ \bibinfo {author}
  {\bibfnamefont {Y.}~\bibnamefont {Zhang}},\ }\bibfield  {title} {\bibinfo
  {title} {Quantum circuits with many photons on a programmable nanophotonic
  chip},\ }\href {https://doi.org/10.1038/s41586-021-03202-1} {\bibfield
  {journal} {\bibinfo  {journal} {Nature}\ }\textbf {\bibinfo {volume} {591}},\
  \bibinfo {pages} {54} (\bibinfo {year} {2021})}\BibitemShut {NoStop}%
\bibitem [{\citenamefont {Nielsen}\ and\ \citenamefont
  {Chuang}(2002)}]{nielsen2002quantum}%
  \BibitemOpen
  \bibfield  {author} {\bibinfo {author} {\bibfnamefont {M.~A.}\ \bibnamefont
  {Nielsen}}\ and\ \bibinfo {author} {\bibfnamefont {I.}~\bibnamefont
  {Chuang}},\ }\href {https://doi.org/10.1017/CBO9780511976667} {\emph
  {\bibinfo {title} {Quantum computation and quantum information}}}\ (\bibinfo
  {publisher} {Cambridge University Press},\ \bibinfo {year}
  {2002})\BibitemShut {NoStop}%
\bibitem [{\citenamefont {Kumar}\ and\ \citenamefont
  {Dhand}(2021)}]{kumar2021unitary}%
  \BibitemOpen
  \bibfield  {author} {\bibinfo {author} {\bibfnamefont {S.~P.}\ \bibnamefont
  {Kumar}}\ and\ \bibinfo {author} {\bibfnamefont {I.}~\bibnamefont {Dhand}},\
  }\bibfield  {title} {\bibinfo {title} {Unitary matrix decompositions for
  optimal and modular linear optics architectures},\ }\href
  {https://doi.org/10.1088/1751-8121/abd4ae} {\bibfield  {journal} {\bibinfo
  {journal} {J. Phys. A}\ }\textbf {\bibinfo {volume} {54}},\ \bibinfo {pages}
  {045301} (\bibinfo {year} {2021})}\BibitemShut {NoStop}%
\bibitem [{\citenamefont {Dhand}\ and\ \citenamefont
  {Goyal}(2015)}]{Dhand2015a}%
  \BibitemOpen
  \bibfield  {author} {\bibinfo {author} {\bibfnamefont {I.}~\bibnamefont
  {Dhand}}\ and\ \bibinfo {author} {\bibfnamefont {S.~K.}\ \bibnamefont
  {Goyal}},\ }\bibfield  {title} {\bibinfo {title} {Realization of arbitrary
  discrete unitary transformations using spatial and internal modes of light},\
  }\href {https://doi.org/10.1103/PhysRevA.92.043813} {\bibfield  {journal}
  {\bibinfo  {journal} {Phys. Rev. A}\ }\textbf {\bibinfo {volume} {92}},\
  \bibinfo {pages} {043813} (\bibinfo {year} {2015})},\ \Eprint
  {https://arxiv.org/abs/1508.06259} {arXiv:1508.06259} \BibitemShut {NoStop}%
\bibitem [{\citenamefont {Su}\ \emph {et~al.}(2019)\citenamefont {Su},
  \citenamefont {Dhand}, \citenamefont {Helt}, \citenamefont {Vernon},\ and\
  \citenamefont {Br\'adler}}]{Su2019a}%
  \BibitemOpen
  \bibfield  {author} {\bibinfo {author} {\bibfnamefont {D.}~\bibnamefont
  {Su}}, \bibinfo {author} {\bibfnamefont {I.}~\bibnamefont {Dhand}}, \bibinfo
  {author} {\bibfnamefont {L.~G.}\ \bibnamefont {Helt}}, \bibinfo {author}
  {\bibfnamefont {Z.}~\bibnamefont {Vernon}},\ and\ \bibinfo {author}
  {\bibfnamefont {K.}~\bibnamefont {Br\'adler}},\ }\bibfield  {title} {\bibinfo
  {title} {Hybrid spatiotemporal architectures for universal linear optics},\
  }\href {https://doi.org/10.1103/PhysRevA.99.062301} {\bibfield  {journal}
  {\bibinfo  {journal} {Phys. Rev. A}\ }\textbf {\bibinfo {volume} {99}},\
  \bibinfo {pages} {062301} (\bibinfo {year} {2019})},\ \Eprint
  {https://arxiv.org/abs/1812.07939} {arXiv:1812.07939} \BibitemShut {NoStop}%
\bibitem [{\citenamefont {Durstenfeld}(1964)}]{Durstenfeld1964}%
  \BibitemOpen
  \bibfield  {author} {\bibinfo {author} {\bibfnamefont {R.}~\bibnamefont
  {Durstenfeld}},\ }\bibfield  {title} {\bibinfo {title} {Algorithm 235: Random
  permutation},\ }\href {https://doi.org/10.1145/364520.364540} {\bibfield
  {journal} {\bibinfo  {journal} {Commun. ACM}\ }\textbf {\bibinfo {volume}
  {7}},\ \bibinfo {pages} {420} (\bibinfo {year} {1964})}\BibitemShut {NoStop}%
\bibitem [{\citenamefont {Okounkov}(2000)}]{Okounkov2000}%
  \BibitemOpen
  \bibfield  {author} {\bibinfo {author} {\bibfnamefont {A.}~\bibnamefont
  {Okounkov}},\ }\bibfield  {title} {\bibinfo {title} {Random matrices and
  random permutations},\ }\href {https://doi.org/10.1155/S1073792800000532}
  {\bibfield  {journal} {\bibinfo  {journal} {Int. Math. Res. Notices}\
  }\textbf {\bibinfo {volume} {2000}},\ \bibinfo {pages} {1043} (\bibinfo
  {year} {2000})}\BibitemShut {NoStop}%
\bibitem [{\citenamefont {Brent}(1971)}]{brent1971algorithm}%
  \BibitemOpen
  \bibfield  {author} {\bibinfo {author} {\bibfnamefont {R.~P.}\ \bibnamefont
  {Brent}},\ }\bibfield  {title} {\bibinfo {title} {An algorithm with
  guaranteed convergence for finding a zero of a function},\ }\href
  {https://doi.org/10.1093/comjnl/14.4.422} {\bibfield  {journal} {\bibinfo
  {journal} {Comput. J.}\ }\textbf {\bibinfo {volume} {14}},\ \bibinfo {pages}
  {422} (\bibinfo {year} {1971})}\BibitemShut {NoStop}%
\end{thebibliography}
\end{document}